\documentclass[twocolumn, graphics, floatfix, a4paper, 10pt, aps, prb, superscriptaddress,showpacs,longbibliography]{revtex4-1}
\usepackage{graphicx}
\usepackage{bm}
\usepackage[usenames,dvipsnames]{xcolor}
\usepackage[colorlinks=true,urlcolor=blue,citecolor=blue,linkcolor=blue]{hyperref}
\usepackage{amssymb,ulem,amsmath}
\usepackage{longtable}
\newcommand{\bra}[1]{\left\langle #1\right|}
\newcommand{\ket}[1]{\left|#1\right\rangle}
\newcommand{\braket}[2]{\left\langle #1 | #2 \right\rangle}
\newcommand{\ketbra}[2]{|#1 \left\rangle\right\langle #2|}
\newcommand{\qql}{\textquotedblleft}
\newcommand{\qqr}{\textquotedblright}
\newcommand{\vc}[1]{\bm{\mathrm{#1}}}

\DeclareMathOperator{\Tr}{Tr}
\DeclareMathOperator{\rk}{rk}
\begin{document}
\title{Effective theory and emergent $SU(2)$ symmetry in the flat bands of attractive Hubbard models}
\author{Murad Tovmasyan}
\affiliation{Institute for Theoretical Physics, ETH Zurich, 8093 Z\"urich, Switzerland}
\author{Sebastiano Peotta}
\affiliation{COMP Centre of Excellence, Department of Applied Physics, Aalto University School of Science, FI-00076 Aalto, Finland}
\author{P\"aivi T\"orm\"a}
\email{ptorma@aalto.fi}
\affiliation{COMP Centre of Excellence, Department of Applied Physics, Aalto University School of Science, FI-00076 Aalto, Finland}
\author{Sebastian D. Huber}
\email{sebastian.huber@phys.ethz.ch}
\affiliation{Institute for Theoretical Physics, ETH Zurich, 8093 Z\"urich, Switzerland}

\begin{abstract}
In a partially filled flat Bloch band electrons do not have a well defined Fermi surface and hence the low-energy theory is not a Fermi liquid. Neverethless, under the influence of an attractive interaction, a superconductor well described by the Bardeen-Cooper-Schrieffer (BCS) wave function can arise. Here we study the low-energy effective Hamiltonian of a generic Hubbard model with a flat band. We obtain an effective Hamiltonian for the flat band physics by eliminating higher lying bands via perturbative Schrieffer-Wolff transformation. At first order in the interaction energy we recover the usual procedure of projecting the interaction term onto the flat band Wannier functions. We show that the BCS wave function is the exact ground state of the projected interaction Hamiltonian and that the compressibility is diverging as a consequence of an emergent $SU(2)$ symmetry. This symmetry is broken by second order interband transitions resulting in a finite compressibility, which we illustrate for a one-dimensional ladder with two perfectly flat bands. These results motivate a further approximation leading to an effective ferromagnetic Heisenberg model. The gauge-invariant result for the superfluid weight of a flat band can be obtained from the ferromagnetic Heisenberg model only if the maximally localized Wannier functions in the Marzari-Vanderbilt sense are used. Finally, we prove an important inequality $D \geq \mathcal{W}^2$ between the Drude weight $D$ and the winding number $\mathcal{W}$, which guarantees ballistic transport for topologically nontrivial flat bands in one dimension.
\end{abstract} 
\maketitle
\section{Introduction}
In quantum mechanics particles can localize due to the destructive interference between different classical trajectories. Such localization can result in the formation of flat bands with a diverging effective mass, or equivalently zero group velocity. Historically the first example of this phenomenon is the formation of the Landau levels of a particle in the presence of a uniform magnetic field. While this localization is a purely single-particle effect, the presence of a flat band can have a profound impact on the physics of interacting many-body systems. A prime example is the fractional quantum Hall effect.~\cite{Tsui:1982} Beyond the physics in a strong magnetic field, flat, or nearly-flat bands are a relatively common occurrence in lattice Hamiltonians where they can be realized by engineering suitable hopping matrix elements.~\cite{Lieb:1989, Mielke:1991, Motruk:2012, Creutz:1999, Tovmasyan:2013, Takayoshi:2013, Sticlet:2014, Biondi:2015, Sun:2011, Tang:2011, Neupert:2011} Given the vanishing group velocity, one could expect an electronic flat band system to be a particularly bad conductor. This is certainly true for the high-temperature phase. However, matters are less clear if a superconductor is formed in such a flat band. Here, we investigate this scenario with a special emphasis on the transport properties.

Why can we expect a superconductor to appear in a flat band to begin with? Let us recall that the BCS superconducting transition temperature scales as $T_c \propto \exp(-1/U D_{\textrm F})$, where $U$ is the interaction strength and $D_{\textrm F}$ is the density of states at Fermi energy. Given that $D_{\textrm F}$ diverges in a flat band system, we can indeed hope for a strong superconducting instability on a flat band.~\cite{Khodel:1990, Volovik:1991, Khodel:1994, Kopnin:2011, Heikkila:2011} This now raises questions regarding the superfluid properties of such a flat-band superconductor.

That superfluid transport can occur in the limit of strictly flat bands has been known from the experiments on the exciton condensate in Quantum Hall bilayers.~\cite{Kellogg:2004, Tutuc:2004, Eisenstein:2014} However, it has been realized recently that even flat bands that emerge from the geometry of the lattice can sustain a large superfluid current in the presence of attractive interactions. This occurs because in the flat band limit the superfluid weight is controlled  by a band structure quantity called the quantum metric which is distinct from, but related to, the Chern number.~\cite{Peotta:2015} 

Despite the recent progress in the study of many-body states in the flat bands of lattice Hamiltonians, there are still many open questions. For example, it has been found in the case of bipartite lattices that the ground state in the presence of an attractive Hubbard interaction is given exactly by the BCS wave function.~\cite{Julku:2016} However, it is quite unclear what are the properties of the normal state above the superconducting transition. 

By definition the normal state is not a Fermi liquid since as the interaction is turned off the system becomes an insulator for any filling of the flat band. Indeed, the picture that the divergence of the effective mass implies the absence of transport is true for a noninteracting system. 

Transport in a flat band is a consequence of either interaction or disorder. It is therefore interesting to characterize the properties both of the normal state and of the superconducting state at nonzero temperature as these can be accessed in current ultracold gas experiments. Moreover, the situation where the superconducting ground state is well-known but the normal metallic state is less understood is reminiscent of the one in high-$T_{\rm c}$ superconductors, especially cuprates and iron-pnictides, where the enigmatic  pseudogap phase is believed to be important for unveiling the pairing mechanism that gives rise to the superconducting phase. It may be possible that idealized flat band models can provide clues in this direction.
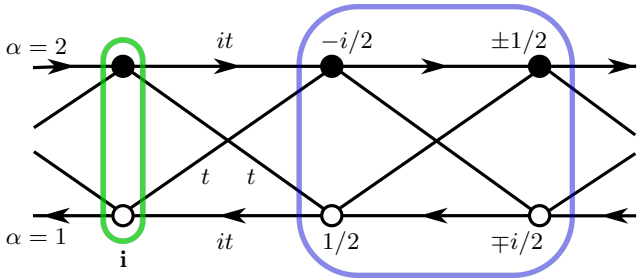
\begin{figure}[tb]
\centering
  \ifx\svgwidth\undefined%
    \setlength{\unitlength}{227.45921691bp}%
    \ifx\svgscale\undefined%
      \relax%
    \else%
      \setlength{\unitlength}{\unitlength * \real{\svgscale}}%
    \fi%
  \else%
    \setlength{\unitlength}{\svgwidth}%
  \fi%
  \global\let\svgwidth\undefined%
  \global\let\svgscale\undefined%
\begin{picture}(1,0.45470552)%
    \put(0,0){\includegraphics{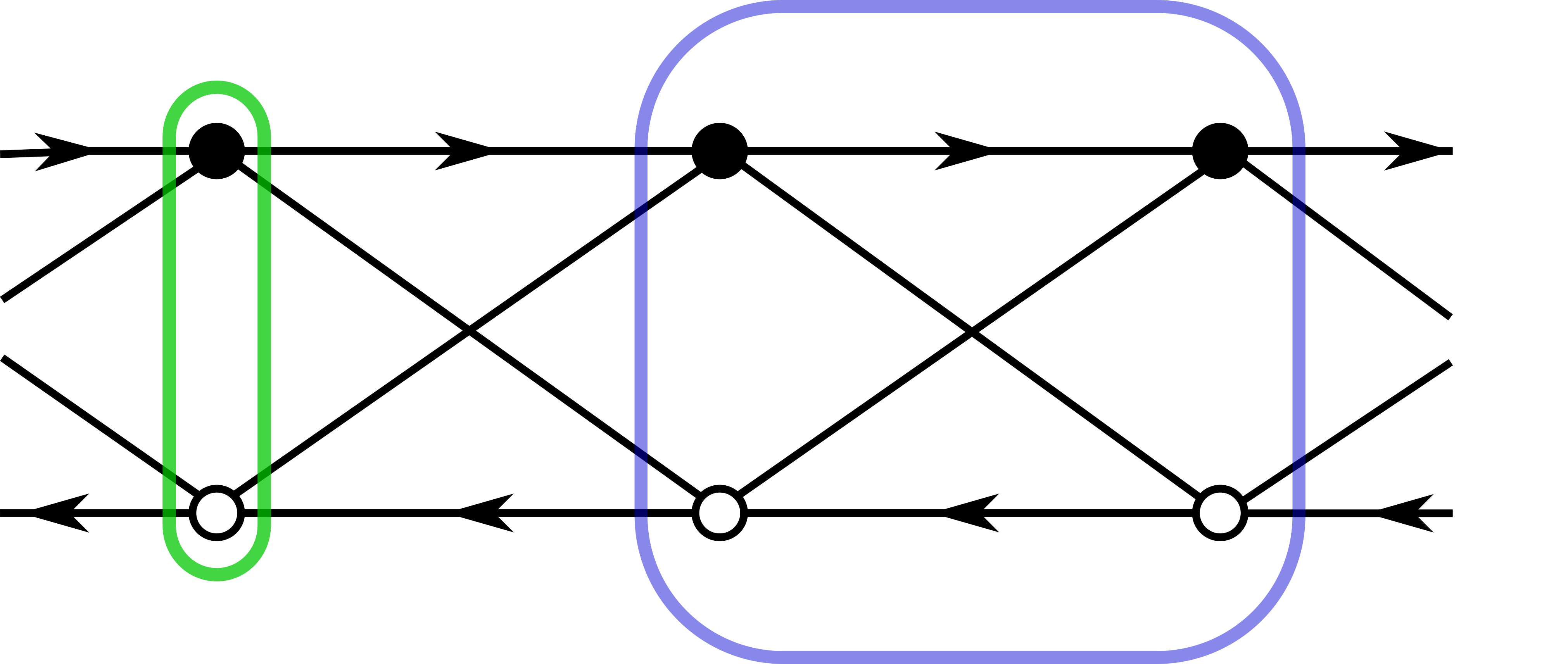}}%
    \put(0.141,0.017){\color[rgb]{0,0,0}\makebox(0,0)[lb]{\smash{$\vc{i}$}}}%
    \put(0.30,0.38){\color[rgb]{0,0,0}\makebox(0,0)[lb]{\smash{$it$}}}%
    \put(0.30,0.047){\color[rgb]{0,0,0}\makebox(0,0)[lb]{\smash{$it$}}}%
    \put(0.275,0.155){\color[rgb]{0,0,0}\makebox(0,0)[lb]{\smash{$t$}}}%
    \put(0.35,0.155){\color[rgb]{0,0,0}\makebox(0,0)[lb]{\smash{$t$}}}%
    \put(0.47098549,0.38){\color[rgb]{0,0,0}\makebox(0,0)[lb]{\smash{$-i/2$}}}%
    \put(0.47372465,0.047){\color[rgb]{0,0,0}\makebox(0,0)[lb]{\smash{$1/2$}}}%
    \put(0.75,0.047){\color[rgb]{0,0,0}\makebox(0,0)[lb]{\smash{$\mp i/2$}}}%
    \put(0.75,0.38){\color[rgb]{0,0,0}\makebox(0,0)[lb]{\smash{$\pm 1/2$}}}%
   \put(-0.045,0.055){\color[rgb]{0,0,0}\makebox(0,0)[lb]{\smash{$\alpha=1$}}}%
   \put(-0.045,0.37){\color[rgb]{0,0,0}\makebox(0,0)[lb]{\smash{$\alpha=2$}}}%
\end{picture}%
\caption{\label{fig:Creutz}(Color online) \textit{The Creutz ladder:} The green box indicates the $\vc{i}$-th unit cell containing two orbitals depicted by an empty and a full circle. The two sublattices are labeled by $\alpha$. The hopping amplitudes for the spin-$\uparrow$ fermions are given on corresponding links. The arrows indicate the sense of the complex hoppings. The blue box indicates the maximally localized Wannier function for the upper/lower ($\pm$) flat band; the numbers correspond to the respective amplitudes which are nonzero only for the sites inside the box.}
\end{figure}  

In this work we address the general problem of providing a reliable low-energy effective theory for the flat band of a multiband lattice Hamiltonian in the presence of an attractive Hubbard interaction. It is important to consider a multiband Hamiltonian since in the case of a single band Hamiltonian, i.e., a Hamiltonian defined on a simple lattice, the only flat band that can be obtained is trivial and corresponds to the limit where all the sites are decoupled (atomic limit). Furthermore, in the multiband case the flat band can have a wide variety of properties encoded in suitable invariants constructed from the Bloch/Wannier functions. An example is the Chern number $\mathcal{C}$, a topological invariant signaling, when nonzero, that the flat band cannot be connected adiabatically to the atomic limit ($\mathcal{C}=0$) or to a band with different Chern number. Interestingly, even a topologically trivial flat band ($\mathcal{C}=0$) can have a nonzero quantum metric and therefore host a superconducting state.~\cite{Peotta:2015} The subject of multiband superconductivity has recently become important with the discovery of materials such as magnesium diboride and iron pnictides superconductors. Flat band superconductivity is an exotic example of the rich variety of phenomena encountered in multiband superconductivity.

The strategy of this work is to combine the result of Ref.~\onlinecite{Peotta:2015} for the superfluid weight of a flat band in terms of the quantum metric with the general approach of Ref.~\onlinecite{Huber:2010} where flat bands are studied by projecting the interaction Hamiltonian on the Wannier functions of the flat band. This latter approach has the advantage of providing a simple low-energy effective Hamiltonian which is often accurate in predicting the properties of the ground state. 

We are able to prove the  useful result that, in the wide class of Hubbard models considered here, the BCS wave function is the exact ground state of the projected interaction Hamiltonian under a simple condition on the Wannier functions. Concomitantly, the compressibility in the partially filled flat band is diverging. In fact we show how both these properties are the manifestation of an emerging $SU(2)$ symmetry, which is due to the band flatness.

Guided by these rigorous results, we are lead to approximate the projected interaction Hamiltonian by an effective ferromagnetic Heisenberg model.
The drawback is that the Wannier functions are defined up to a unitary transformation, and therefore any approximation performed on the projected interaction Hamiltonian depends on the specific choice of the Wannier functions. Using the gauge invariant result of Ref.~\onlinecite{Peotta:2015}, we show that there is a preferred choice for the Wannier functions, which coincides with the maximally localized Wannier functions in the Marzari-Vanderbilt sense. 

As a concrete example of our general results, we consider an Hubbard model defined on a one-dimensional ladder, the Creutz ladder, which is graphically defined in Fig.~\ref{fig:Creutz}. The band structure of the Creutz ladder consists of two perfectly flat bands. In the case of the attractive Creutz-Hubbard model the resulting effective ferromagnetic Heisenberg model takes the form of the integrable $XXX$ chain. 

We find also that in order to account for a finite compressibility it is necessary to include the effect of interband transitions resulting in higher order terms in the effective Hamiltonian. We provide an analytic result for the compressibility up to second order in the ratio $U/E_{\rm gap}$ between interaction and band gap in the case of the Creutz-Hubbard model. This result is tested against Density Matrix Renormalization Group (DMRG) simulations and we find an excellent agreement.

Finally, we prove an important bound $D \geq \mathcal{W}^2$ between Drude weight $D$  and the one-dimensional winding number, which extends the result for the superfluid weight $D_{\rm s} \geq |\mathcal{C}|$ valid in two dimensions\cite{Peotta:2015}. It is shown using the Creutz-Hubbard model that the inequality is in fact optimal. 

The paper is organized as follows. In Section~\ref{sec:SW} we introduce the model, the basic notations, and the perturbative Schrieffer-Wolff (SW) transformation for a generic Hubbard model. In Section~\ref{sec:exact_BCS} we derive the result that the BCS wave function is exact in the isolated flat band limit and provide the generators of the emergent $SU(2)$ symmetry. In Section~\ref{sec:spin_Ham} we derive the ferromagnetic Heisenberg model by dropping the pair-breaking terms in the projected interaction Hamiltonian. We show that this mapping is in fact exact for the Creutz-Hubbard model. In Section~\ref{sec:SFW_spin} we discuss how the result of Ref.~\onlinecite{Peotta:2015} for the superfluid weight of a flat band can be recovered from the Heisenberg model. This is done by introducing an overlap functional for Wannier functions, whose relation with the usual Marzari-Vanderbilt functional is analyzed in detail. The proof of various results relating the two functionals is detailed in Appendix~\ref{app:functional_relations}. In Section~\ref{sec:second_order} we derive the second order corrections to the effective Hamiltonian using the SW transformation in the case of the Creutz-Hubbard model. In Section~\ref{sec:compressibility} we use the result of the previous section to derive an analytic result for the compressibility of the Creutz-Hubbard model which is then compared to DMRG simulations. In Section~\ref{sec:winding} the inequality between superfluid weight and winding number is proved. Our results and the future perspectives are discussed in the last Section.

\section{Effective Hamiltonian from the SW transformation}
\label{sec:SW}
\subsection{The model}\label{subsec:model}
Here we introduce the class of flat band models with attractive Hubbard interactions defined on a $d$-dimensional lattice that are considered in this work. We shall also give some useful definitions and consider an one-dimensional example, the Creutz-Hubbard model.

The kinetic part of the tight-binding Hamiltonian $\hat{\mathcal{H}}=\hat{\mathcal{H}}_{\rm kin}+\hat{\mathcal{H}}_{\rm int}$ is given by
\begin{equation}\label{eq:Ham_kin}
\mathcal{\hat{H}}_{\rm kin} =\sum_{\sigma=\uparrow,\downarrow}\, \sum_{\vc{i}\alpha,\vc{j}\beta}t^{\sigma}_{\vc{i}\alpha,\vc{j}\beta}\,\hat{c}^\dagger_{\vc{i}\alpha\sigma}\hat{c}_{\vc{j}\beta\sigma}\,,
\end{equation}
where, as usual, $\hat{c}^\dagger_{\vc{i}\alpha\sigma}$ and $\hat{c}_{\vc{i}\alpha\sigma}$ are the creation and the annihilation operators, respectively, of a fermion with spin $\sigma$ at unit cell $\vc{i}$ and sublattice $\alpha$. According to Eq.~\eqref{eq:Ham_kin} the number of spin-$\uparrow$ and spin-$\downarrow$ particles are separately conserved. In order to preserve the time-reversal symmetry and favor the occurrence of Cooper pairs in the spirit of Anderson's theorems,~\cite{Anderson:1959,Anderson:1984} we take the hopping matrix $(t^{\sigma}_{\vc{i}\alpha,\vc{j}\beta})$ such that the kinetic Hamiltonian is time-reversal invariant, namely $t^{\uparrow}_{\vc{i}\alpha,\vc{j}\beta}=(t^{\downarrow}_{\vc{i}\alpha,\vc{j}\beta})^*$, where the star denotes the complex conjugate. Moreover, we consider models in which the kinetic Hamiltonian has an \textit{isolated flat Bloch band} separated from the other bands by a finite energy gap $E_{\rm gap}$. Without loss of generality, we concentrate on the case where the flat band is the lowest lying band to simplify the notation.

The attractive Hubbard interaction term has the form 
\begin{equation}\label{eq:int_Ham}
\mathcal{\hat{H}}_{\rm int} = -U\sum_{\vc{i}\alpha}\hat{n}_{\vc{i}\alpha\uparrow}
\hat{n}_{\vc{i}\alpha\downarrow}\,,
\end{equation}
where $\hat{n}_{\vc{i}\alpha\sigma} =\hat{c}^\dagger_{\vc{i}\alpha\sigma}\hat{c}_{\vc{i}\alpha\sigma}$ is the fermionic number operator and we assume that $U>0$. 

Occasionally we comment on the corresponding repulsive Hubbard model, where by \qql corresponding\qqr we mean that the spin-$\uparrow$ and spin-$\downarrow$ kinetic Hamiltonians are equal ($t^{\uparrow}_{\vc{i}\alpha,\vc{j}\beta}=t^{\downarrow}_{\vc{i}\alpha,\vc{j}\beta}$) and equal to the spin-$\uparrow$ kinetic Hamiltonian of the attractive model. Thus the corresponding Hubbard model possess full $SU(2)$ spin rotational symmetry. We will see that there is a general duality between flat band superconductivity in the attractive model and flat band ferromagnetism in the corresponding repulsive model.

For the discussion of the interacting problem below, the Wannier states form a convenient local orthonormal basis.~\cite{Huber:2010,Tovmasyan:2013} The Wannier function of the flat band centered at the unit cell $\vc{j}$ is constructed from the Bloch functions $g_{\vc{k}\sigma}(\alpha)$ of the flat band according to~\cite{Wannier:1937,Marzari:2012}
\begin{equation}\label{eq:Wannier_def}
W_{\alpha\sigma}(\vc{i}-\vc{j}) = 
\frac{V_{\rm c}}{(2\pi)^d}\int_{\rm B.Z.} d^d\vc{k}\,
e^{i\vc{k}\cdot(\vc{r}_{\vc{i}}-\vc{r}_{\vc{j}})} g_{\vc{k}\sigma}(\alpha)\,,
\end{equation}
where $V_{\rm c}$ denotes the volume of the unit cell and $\vc{r}_{\vc{i}}$ is the lattice vector corresponding to unit cell $\vc{i}$.  

It is a well known fact  that the Wannier functions are not uniquely defined because of the gauge freedom of the Bloch functions. In fact, we can change the Bloch functions by $g_{\vc{k}\sigma}(\alpha)\rightarrow\exp(i\varphi_{\vc{k}})\,g_{\vc{k}\sigma}(\alpha)$, where $\varphi_{\vc{k}}$ is an arbitrary real function of quasi-momentum $\vc{k}$. Below, when necessary, we discuss the consequences of this gauge freedom in details.

As an example we shall consider the Creutz ladder, a one-dimensional model of two cross-linked chains, depicted in Fig.~\ref{fig:Creutz}. The hopping matrix elements for the spin-$\uparrow$ fermions are also given in Fig.~\ref{fig:Creutz}. The band structure of this model is extremely simple consisting in two perfectly flat bands at $\pm 2t$ ($E_{\rm gap} = 4t$), which allows to evaluate the interband effects analytically. Moreover, the Wannier functions of the Creutz model for both the upper and lower flat band can be chosen to be perfectly localized on only two adjacent rungs, also shown in Fig.~\ref{fig:Creutz}. The lower band Wannier function for the spin-$\uparrow$ fermions correspond to the following periodic and analytic Bloch functions 
\begin{equation}\label{eq:plaquette_Bloch}
g_{k\uparrow}(\alpha) = 
\begin{cases}
e^{\frac{ika}{2}}\sin\left(\frac{ka}{2}+\frac{\pi}{4}\right) & \text {for} \quad\alpha = 1\,,\\
e^{\frac{ika}{2}}\cos\left(\frac{ka}{2}+\frac{\pi}{4}\right) &
\text {for} \quad\alpha = 2\,, \\
\end{cases}
\end{equation}
where $a$ is the lattice spacing. Note that, from the time-reversal invariance of the kinetic Hamiltonian, we have $g_{\vc{k}\downarrow}(\alpha)=\left[g_{-\vc{k}\uparrow}(\alpha)\right]^*$ or, equivalently, $W_{\alpha\downarrow}(\vc{i})=[W_{\alpha\uparrow}(\vc{i})]^*$.
\subsection{The effective Hamiltonian} 
In this subsection we give the framework which we use to construct an effective low-energy theory for the class of Hubbard models introduced above. We are interested mainly in the attractive case, but this framework can be used in the case of repulsive Hubbard interactions as well.

A simple low-energy theory can be constructed by projecting the interaction term into the subspace where only the flat band states are occupied. This is done in practice by truncating the expansion of field operators in terms of Wannier orbitals of all bands to retain only the orbitals corresponding to the chosen flat band,~\cite{Huber:2010} thereby restricting the Hilbert space to the flat band subspace. However, one expects interband effects which become more relevant with increasing interaction strength $U$. To include these effects, we use the perturbative SW transformation that allows to take them into account by means of a low-energy effective Hamiltonian, which involves only the degrees of freedom of the lowest flat band.~\cite{Bravyi:2011}

Let us define the field operators projected in the flat band as
\begin{equation}\label{eq:lowerprojected_c}
	\bar{c}_{\vc{i}\alpha\sigma} = \sum_{\vc{j}} W_{\alpha\sigma}(\vc{i}-\vc{j})\,\hat{d}_{\vc{j}\sigma}\,,
\end{equation}
where $W_{\alpha\sigma}(\vc{i}-\vc{j})$ is the Wannier function of the flat band centered at unit cell $\vc{j}$ and $\hat{d}_{\vc{j}\sigma}$ is the annihilation operator corresponding to this Wannier orbital. Hereafter we drop the spin index of the Wannier and Bloch functions and refer to the ones for the spin-$\uparrow$, that is, $W_{\alpha}(\vc{i})\equiv W_{\alpha\uparrow}(\vc{i})$. By virtue of the orthonormality of the Wannier functions, namely $\sum_{\vc{i}\alpha}\left[W_{\alpha}(\vc{i}-\vc{j})\right]^*W_{\alpha}(\vc{i}-\vc{l})= \delta_{\vc{j},\vc{l}}\,$, fermionic operators $\hat{d}_{\vc{i}\sigma},\,\hat{d}_{\vc{i}\sigma}^\dagger$ satisfy the canonical anticommutation relations. It is important to note that the projected field operators satisfy modified anticommutation relations
\begin{gather}\label{eq:anticoms}
\{\bar{c}_{\vc{i}\alpha\uparrow},\bar{c}^\dagger_{\vc{j}\beta\uparrow}\} = P_{\alpha,\beta}(\vc{i}-\vc{j})\,, \quad \{\bar{c}_{\vc{i}\alpha\downarrow},\bar{c}^\dagger_{\vc{j}\beta\downarrow}\} = [P_{\alpha,\beta}(\vc{i}-\vc{j})]^*\,,
\end{gather} 
where we have introduced the lower band projector defined by $P_{\alpha,\beta}(\vc{i}-\vc{j})= \sum_{\vc{l}}W_{\alpha}(\vc{i}-\vc{l})\left[W_{\beta}(\vc{j}-\vc{l})\right]^*$. All other commutation relations are trivial. 

Despite the gauge freedom in the definition of the Wannier functions, the projected field operators given in Eq.~\eqref{eq:lowerprojected_c} are gauge independent, i.e., they are the same for any choice of the Wannier function. Indeed, an equivalent way of defining the projected operators is via the lower band projector introduced above according to $\bar{c}_{\vc{i}\alpha\uparrow} =\sum_{\vc{j}\beta}P_{\alpha,\beta}(\vc{i}-\vc{j})\,\hat{c}_{\vc{j}\beta\uparrow}$ and $\bar{c}_{\vc{i}\alpha\downarrow} =\sum_{\vc{j}\beta}[P_{\alpha,\beta}(\vc{i}-\vc{j})]^*\,\hat{c}_{\vc{j}\beta\downarrow}$. From the last relations the gauge invariance of the projected operators becomes explicit, as the projectors are gauge independent.    

From the projected operator given in Eq.~\eqref{eq:lowerprojected_c} we can define the projected number operator as $\bar{n}_{\vc{i}\alpha\sigma} =\bar{c}_{\vc{i}\alpha\sigma}^\dagger \bar{c}_{\vc{i}\alpha\sigma}$. Accordingly, the projected interaction Hamiltonian becomes $\overline{\mathcal{H}}_{\rm int} = -U\sum_{\vc{i}\alpha}\bar{n}_{\vc{i}\alpha\uparrow}\bar{n}_{\vc{i}\alpha\downarrow}$. This low-energy effective Hamiltonian neglects all interband effects and it describes the system quite well if the minimum band gap $E_{\rm gap}$ between the flat band and other bands is much bigger then the interaction strength $U\ll E_{\rm gap}$. If this condition is not satisfied, one needs to  include the effects from other bands. 

Let us denote the second-quantized projection operator that projects on the subspace of the Hilbert-Fock space where only the states of the flat band have nonzero occupancy by $\mathcal{\hat{P}}=\mathcal{\hat{P}}_\uparrow\mathcal{\hat{P}}_\downarrow$, where $\mathcal{\hat{P}}_\sigma$ is the projector relative to spin $\sigma$. Then the following properties are immediate from the definitions
\begin{equation}
\mathcal{\hat{P}}_\sigma \tilde{c}_{\vc{i}\alpha\sigma}^\dagger = \tilde{c}_{\vc{i}\alpha\sigma}\mathcal{\hat{P}}_\sigma = 0\,,
\end{equation}
where $\tilde{c}_{\vc{i}\alpha\sigma}$ is the field operator projected into the complement of the flat band subspace, namely $\tilde{c}_{\vc{i}\alpha\sigma} = \hat{c}_{\vc{i}\alpha\sigma}- \bar{c}_{\vc{i}\alpha\sigma}$. The operator $\tilde{c}_{\vc{i}\alpha\sigma}$ can be expressed by the Wannier orbitals of the higher bands similarly to~(\ref{eq:lowerprojected_c}). The operators $\mathcal{\hat{P}_\sigma}$ commute with the flat band field operators $\bar{c}_{\vc{i}\alpha\sigma},\,\bar{c}_{\vc{i}\alpha\sigma}^\dagger$ and with the operators with opposite spin. It is then easy to verify that $\mathcal{\hat{P}} \hat{n}_{\vc{i}\alpha\uparrow}\hat{n}_{\vc{i}\alpha\downarrow} \mathcal{\hat{P}} =\mathcal{\hat{P}_\uparrow}\hat{n}_{\vc{i}\alpha\uparrow}\mathcal{\hat{P}_\uparrow}\mathcal{\hat{P}_\downarrow}\hat{n}_{\vc{i}\alpha\downarrow}\mathcal{\hat{P}_\downarrow}=\mathcal{\hat{P}}\bar{n}_{\vc{i}\alpha\uparrow}\bar{n}_{\vc{i}\alpha\downarrow}\mathcal{\hat{P}}$, namely the projected interaction operator $\overline{\mathcal{H}}_{\rm int}$ is obtained by sandwiching $\mathcal{\hat H}_{\rm int}$ with $\mathcal{\hat P}$. We call $\mathcal{\hat Q} = \bm{1}-\mathcal{\hat P}$ the complementary projector.
Using the notation of Ref.~\onlinecite{Bravyi:2011}, we define superoperators $\mathcal{D}(\cdot)$ and $\mathcal{O}(\cdot)$ acting on a generic operator $\hat{X}$ as
\begin{equation}\label{eq:off-diag-def}
\mathcal{D}(\hat{X}) = \mathcal{\hat{P}}\hat{X}\mathcal{\hat{P}} + \mathcal{\hat{Q}}\hat{X}\mathcal{\hat{Q}}\,,\qquad
\mathcal{O}(\hat{X}) = \mathcal{\hat{P}}\hat{X}\mathcal{\hat{Q}} + \mathcal{\hat{Q}}\hat{X}\mathcal{\hat{P}}\,. 
\end{equation}
The superoperator $\mathcal{D}(\cdot)$ extracts the diagonal part of the operator in the argument, while $\mathcal{O}(\cdot)$ the off-diagonal one. Therefore, $\hat{X} = \mathcal{D}(\hat{X}) + \mathcal{O}(\hat{X})$.
Let us define another superoperator $\mathcal{L}(\cdot)$
\begin{equation}
\mathcal{L}(\hat X) = \sum_{i,j} \frac{\ketbra{i}{i} \mathcal{O}(\hat X)\ketbra{j}{j}}{E_i-E_j}\,,
\end{equation}
where the labels $i,j$ run over the eigenstates $E_i,E_j$ of the noninteracting Hamiltonian $\mathcal{\hat{H}}_{\rm kin}$. Since the matrix element $\bra{i}\mathcal{O}(\hat X)\ket{j}$ is nonzero only if $\ket{i}$ belongs to the flat band subspace and $\ket{j}$ to the complementary subspace, or vice versa, the above sum is well defined. It follows that $\mathcal{L}([\mathcal{\hat{H}}_{\rm kin},\hat{X}]) = [\mathcal{\hat{H}}_{\rm kin},\mathcal{L}(\hat{X})] = \mathcal{O}(\hat{X})$. 

With the help of the above introduced superoperators, the effective Hamiltonian up to second order in an expansion in the coupling constant $U$ reads~\cite{Bravyi:2011}
\begin{equation}\label{eq:SW_exp}
\mathcal{\hat H}_{\rm eff} \approx \mathcal{\hat H}_{\rm kin}\mathcal{\hat P} + 
\mathcal{\hat P} \mathcal{\hat H}_{\rm int}\mathcal{\hat P} + \frac{1}{2}\mathcal{\hat P}[\mathcal{L}(\mathcal{\hat H}_{\rm int}),\mathcal{O}(\mathcal{\hat H}_{\rm int})]\mathcal{\hat P}\,.
\end{equation}
The zero order term is the projected kinetic Hamiltonian which is a trivial constant for a flat band, while the first order term is simply the projected interaction term $\mathcal{\overline{H}}_{\rm int}=\mathcal{\hat P} \mathcal{\hat H}_{\rm int}\mathcal{\hat P}=-U\sum_{\vc{i}\alpha}\bar{n}_{\vc{i}\alpha\uparrow}\bar{n}_{\vc{i}\alpha\downarrow}$. Below we will use this general result for the Creutz-Hubbard model. 

\section{Exactness of the BCS wave function and emergent $SU(2)$ symmetry}
\label{sec:exact_BCS}

In this Section we prove that the ground state of the projected attractive interaction $\mathcal{\overline{H}}_{\rm int}$ for arbitrary filling the flat band is given exactly by the BCS wave function. This statement is analogous to the well known fact that the completely polarized ferromagnetic state is the ground state for a half-filled flat band in a repulsive Hubbard model  if the flat band is the lowest lying one\cite{Mielke:1993}. An important difference is that in the former case we known the ground state only of the \textit{projected} Hamiltonian $\mathcal{\overline{H}}_{\rm int}$, while the latter is a statement regarding the \textit{full} Hamiltonian $\mathcal{\hat H}$.

This result generalizes the one relative to bipartite lattices. In a bipartite lattice it is possible to relate a repulsive Hubbard model to an attractive Hubbard model by a particle-hole transformation. Using this transformation and the Lieb theorem\cite{Lieb:1989}, it was shown in Ref.~\onlinecite{Julku:2016} that the completely polarized ferromagnetic state maps into the BCS wave function, which is then the exact ground state of the attractive Hubbard model for arbitrary filling of the flat band. Below we show that the BCS wave function is the exact ground state in a more general setting.

First, let us define the operator $\hat{b}_0^\dagger$ which creates a Cooper pair in a plane wave state with zero quasi-momentum in the flat band
\begin{equation}\label{eq:pair_creat}
\hat{b}_0^\dagger = \sum_{\vc{k}}\hat{d}_{\vc{k}\uparrow}^\dagger\hat{d}_{-\vc{k}\downarrow}^\dagger= \sum_{\vc{j}}\hat{d}_{\vc{j}\uparrow}^\dagger
\hat{d}_{\vc{j}\downarrow}^\dagger\,.
\end{equation} 
The above introduced operator $\hat{d}_{\vc{k}\sigma}^\dagger$ creates a fermion with quasi-momentum $\vc{k}$ in the flat Bloch band. Note that the pair creation operator takes the above  simple form only in the flat band limit. Generally, it is given by $\hat{b}_0^\dagger =\sum_{\vc{k}}g_{\vc{k}}\,\hat{d}_{\vc{k}\uparrow}^\dagger\hat{d}_{-\vc{k}\downarrow}^\dagger$, where $g_{\vc{k}}= v_{\vc{k}}/u_{\vc{k}}$ defines the Cooper pair wave function. In the flat band limit the BCS coherence factors $u_{\vc{k}} = u = \sqrt{1-\nu}$ and $v_{\vc{k}} = v =\sqrt{\nu}$  are independent of the quasi-momentum $\vc{k}$ and only in this case the second equality in Eq.~\eqref{eq:pair_creat} is valid. Here $\nu=N/(2N_{\rm c})$ is the filling of the flat band with $N_{\rm c}$ the number of unit cells and $N$ the number of fermions. Hence, the BCS wave function in the grand canonical ensemble can be written in the following equivalent forms
\begin{equation}\label{eq:BCS_grand}
\begin{split}
\ket{\Omega} = u^{N_{\rm c}}\exp\left(\frac{v}{u}\,\hat{b}_0^\dagger\right) \ket{\emptyset} &= \prod_{\vc{j}} \left(u+v\hat{d}_{\vc{j}\uparrow}^\dagger\hat{d}_{\vc{j}\downarrow}^\dagger\right)\ket{\emptyset}\\
& = \prod_{\vc{k}} \left(u+v\hat{d}_{\vc{k}\uparrow}^\dagger\hat{d}_{-\vc{k}\downarrow}^\dagger\right)\ket{\emptyset}\,,
\end{split}
\end{equation}
where $\ket{\emptyset}$ is the vacuum containing no fermions. In expanding the exponential in Eq.~(\ref{eq:BCS_grand}) we have made use of Fermi statistics, which imply $\big(\hat{d}_{\vc{j}\sigma}^\dagger\big)^2=\big(\hat{d}_{\vc{k}\sigma}^\dagger\big)^2=0$.

It is straightforward to check that $\hat{b}_0^\dagger$ commutes with the \textit{projected} spin operator $\overline{S}^{z}_{\vc{i}\alpha}=\left(\bar{n}_{\vc{i}\alpha\uparrow}-\bar{n}_{\vc{i}\alpha\downarrow}\right)/2$, i.e., $[\hat{b}_0^{\dagger}\,,\,\overline{S}^{z}_{\vc{i}\alpha}]=0$. From this commutation relation one has that $\overline{S}^{z}_{\vc{i}\alpha}\ket{\Omega} = 0$.  Therefore, the BCS wave function $\ket{\Omega}$ is a zero eigenvector of the positive semidefinite operator $\mathcal{\overline{H}}_{\rm int}'= (U/2)\sum_{\vc{i}\alpha}(\bar{n}_{\vc{i}\alpha\uparrow}-\bar{n}_{\vc{i}\alpha\downarrow})^2$. Note that, as a consequence of the modified anticommutation relations~(\ref{eq:anticoms}), $\bar{n}_{\vc{i}\alpha\sigma}^2 = P_{\alpha\alpha}(\vc{0})\bar{n}_{\vc{i}\alpha\sigma}$, while the usual number operators satisfy $\hat{n}_{\vc{i}\alpha\sigma}^2 =\hat{n}_{\vc{i}\alpha\sigma}$. Using this, one can expand the square
\begin{equation}
\mathcal{\overline{H}}_{\rm int}' = \frac{U}{2}\sum_{\vc{i}\alpha}
P_{\alpha\alpha}(\vc{0})
\left(\bar{n}_{\vc{i}\alpha\uparrow}+
\bar{n}_{\vc{i}\alpha\downarrow}\right)
-U\sum_{\vc{i}\alpha}\bar{n}_{\vc{i}\alpha\uparrow}
\bar{n}_{\vc{i}\alpha\downarrow}\,.
\end{equation}
The last term in the above equation is precisely the attractive Hubbard interaction, while the first term is in general a nontrivial orbital-resolved potential. Consider now the case in which the flat band Bloch/Wannier functions have the same weight on the orbitals were they are nonzero, which means
\begin{equation}
\label{eq:condition_exact}
\begin{split}
n_\phi &= P_{\alpha\alpha}(\vc{0})  = P_{\beta\beta}(\vc{0}) \\
&= \sum_{\vc{i}}|W_{\alpha}(\vc{i})|^2 = \frac{V_{\rm c}}{(2\pi)^d}\int_{\rm B.Z.} d^d\vc{k}\,|g_{\vc{k}}(\alpha)|^2\,,
\end{split}
\end{equation}
in a certain subset  $\alpha,\beta \in \mathcal{S}$ of orbitals and $P_{\gamma\gamma}(\vc{0})=0$ for $\gamma \not\in \mathcal{S}$.
Here $n_{\phi}^{-1} = |\mathcal{S}|$ is the number of orbitals on which the flat band states have nonvanishing weight and, equivalently, on which the Cooper pair wave function is uniformly delocalized. We call the Eq.~\eqref{eq:condition_exact} the uniform pairing condition.  Then the operator $\mathcal{\overline{H}}_{\rm int}'$ reduces to
\begin{equation}\label{eq:N-H}
\mathcal{\overline{H}}_{\rm int}' = \frac{n_\phi U}{2}\overline{N}
-U\sum_{\vc{i}\alpha}\bar{n}_{\vc{i}\alpha\uparrow}
\bar{n}_{\vc{i}\alpha\downarrow}\,,
\end{equation}
with $\overline{N} = \sum_{\vc{j}\sigma}\hat{d}_{\vc{j}\sigma}^\dagger\hat{d}_{\vc{j}\sigma}$ the projected particle number operator. In this case $\mathcal{\overline{H}}_{\rm int}'$ differs from the projected attractive Hubbard interaction $\mathcal{\overline{H}}_{\rm int}$ by a trivial term proportional to the  particle number operator. Following the usual argument~\cite{Mielke:1993} we conclude that the BCS wave function is the ground state of $\mathcal{\overline{H}}_{\rm int}$ if the condition in Eq.~\eqref{eq:condition_exact} is satisfied. The ground state energy is 
\begin{equation}\label{eq:E_BCS}
\frac{E_{\rm BCS}}{N_{\rm c}} = (2\varepsilon_0-n_\phi U)\nu
\end{equation}
with $\varepsilon_0$ the flat band energy. It is important to note that this result is only valid asymptotically for small $U\ll E_{\rm gap}$, contrary to the repulsive case. The reason is that the pair creation operator $\hat{b}_0^\dagger$ commutes only with the projected spin operator $\overline{S}^z_{\vc{i}\alpha}$ and it does not commute with the full spin operator $\hat{S}^z_{\vc{i}\alpha}$. In fact one can show that $[\hat{S}^{z}_{\vc{i}\alpha},\hat{b}_0^\dagger]=\frac{1}{2}(\tilde{c}_{\vc{i}\alpha\uparrow}^\dagger\bar{c}_{\vc{i}\alpha\downarrow}^\dagger-\bar{c}_{\vc{i}\alpha\uparrow}^\dagger\tilde{c}_{\vc{i}\alpha\downarrow}^\dagger)$.

From the Cooper pair creation operator $\hat{b}_0^{\dagger}\,$, we can construct the generators of the emergent $SU(2)$ symmetry of the projected interaction Hamiltonian which also includes the usual $U(1)$ particle number conservation symmetry. The operators $(1/2)(\hat{b}^{\dagger}_0+\hat{b}_0)$, $(-i/2)(\hat{b}^{\dagger}_0-\hat{b}_0)$ and $(1/2)[\hat{b}_0^{\dagger},\hat{b}_0]$ form the generators of $SU(2)$. Using the commutator $[b_0^{\dagger},[\hat{b}_0^{\dagger},\hat{b}_0]]=-2b_0^{\dagger}\,$, it is straightforward to check that the three generators given above satisfy the commutation relations of the $\mathfrak{su}(2)$ algebra. 

A consequence of this symmetry is that the compressibility is diverging; as it also follows from Eq.~\eqref{eq:E_BCS} and the definition of the inverse compressibility $\kappa^{-1} = \nu^2\partial^2(E_{\rm BCS}/N_{\rm c})/\partial\nu^2$.
Moreover, since $\ket{\Omega}$ is a zero eigenstate of $\mathcal{\overline{H}}_{\rm int}'$, one obtains the relation  $n_\phi\langle \overline{n}_{\vc{i}\alpha\uparrow} \rangle = n_\phi\langle\overline{n}_{\vc{i}\alpha\downarrow} \rangle= \langle \overline{n}_{\vc{i}\alpha\uparrow}
\overline{n}_{\vc{i}\alpha\downarrow}\rangle$ between density and double occupancy (expectation values are here taken on $\ket{\Omega}$). Both the results for the energy and the relation between density and double occupancy have been verified with DMRG for the Creutz-Hubbard model ($n_\phi = 1/2$).

We mention also that, as a consequence of the uniform pairing condition~\eqref{eq:condition_exact}, the expectation values of the projected density $\langle \bar{n}_{\alpha\sigma}\rangle$ and of the pairing order parameters $\Delta_{\alpha}=-U \langle \hat{c}_{\vc{i}\alpha\downarrow}\hat{c}_{\vc{i}\alpha\uparrow}\rangle$ are constant as a function of $\alpha\in \mathcal{S}$ and vanish for $\alpha \not\in \mathcal{S}$. In particular, the condition of constant $\Delta_{\vc{i}\alpha}$ has been used in Ref.~\onlinecite{Peotta:2015} to derive the relation between superfluid weight and quantum metric.

\section{Spin chain form of the effective Hamiltonian}
\label{sec:spin_Ham}

As a next step, we show that it is possible to drop a large amount of terms in the projected interaction Hamiltonian $\mathcal{\overline{H}}_{\rm int}$ at the same time preserving all of the properties mentioned above, namely the BCS ground state and the emergent $SU(2)$ symmetry. The result of this truncation is an effective ferromagnetic Heisenberg model, where the components of the pseudospin are the creation, annihilation and occupation operator of a Cooper pair in a Wannier function. The effective spin model is computationally much easier to deal with and offers an intuitive model of a flat band superconductor as a ferromagnet.

The effective Hamiltonian obtained by the projection technique with Wannier functions  is generally given by 
\begin{equation}\label{eq:expansion_projH}
\overline{\mathcal{H}}_{\rm int}=-U\sum_{\vc{i},\vc{j},\vc{k},\vc{l}}T_{\vc{i}\vc{j}\vc{k}\vc{l}}\,\hat{d}^{\dagger}_{\vc{i}\uparrow}\hat{d}^{\dagger}_{\vc{j}\downarrow}\hat{d}_{\vc{k}\downarrow}\hat{d}_{\vc{l}\uparrow}\,,
\end{equation}
where the coefficients $T_{\vc{i}\vc{j}\vc{k}\vc{l}}$ can be written in terms of Wannier functions. In the case of attractive interactions it is energetically more favorable that fermions with opposite spins form a Cooper pair. Hence, we truncate the projected Hamiltonian to the subspace defined by the conditions $\big(\hat{d}_{\vc{i}\uparrow}^\dagger
\hat{d}_{\vc{i}\uparrow} -  
\hat{d}_{\vc{i}\downarrow}^\dagger
\hat{d}_{\vc{i}\downarrow}\big)\ket{\psi} = 0$ for every $\vc{i}$. In other words, only the terms in the expansion~\eqref{eq:expansion_projH} that preserve and act nontrivially in this subspace are retained. After this truncation, the only remaining degree of freedom is the presence or absence of a Cooper pair in a given Wannier state. This is encoded in the components of the pseudospin $\hat{\vc{S}}_{\vc{i}}$ defined by
\begin{equation}\label{eq:def_spin2}
\hat{\mathrm{S}}^z_{\vc{i}} = \frac{1}{2}(\hat{d}_{\vc{i}\uparrow}^\dagger
\hat{d}_{\vc{i}\uparrow} + 
\hat{d}_{\vc{i}\downarrow}^\dagger
\hat{d}_{\vc{i}\downarrow}-1),\quad 
\hat{\mathrm{S}}^+_{\vc{i}} = \hat{d}_{\vc{i}\uparrow}^\dagger
\hat{d}_{\vc{i}\downarrow}^\dagger,\quad
\hat{\mathrm{S}}^-_{\vc{i}} = \hat{d}_{\vc{i}\downarrow}
\hat{d}_{\vc{i}\uparrow}. 
\end{equation} 
Note that $\hat{b}^{\dagger}_0=\sum_{\vc{i}}\hat{\mathrm{S}}^{+}_{\vc{i}}$ and the generators of the $SU(2)$ symmetry given in the previous section correspond to $\sum_{\vc{i}}\hat{\mathrm{S}}^{x}_{\vc{i}}$, $\sum_{\vc{i}}\hat{\mathrm{S}}^{y}_{\vc{i}}$ and $\sum_{\vc{i}}\hat{\mathrm{S}}^{z}_{\vc{i}}$, where $\hat{\mathrm{S}}_{\vc{i}}^{x}=(1/2)(\hat{\mathrm{S}}^{+}_{\vc{i}}+\hat{\mathrm{S}}^{-}_{\vc{i}})$ and $\hat{\mathrm{S}}_{\vc{i}}^{y}=(-i/2)(\hat{\mathrm{S}}^{+}_{\vc{i}}-\,\hat{\mathrm{S}}^{-}_{\vc{i}})$.  

Using Eq.~\eqref{eq:def_spin2} we can map the projected Hamiltonian (after the truncation) into an effective spin model which is an isotropic ferromagnetic Heisenberg model given by
\begin{equation}\label{eq:eff_Ham}
\mathcal{\hat H}_{\rm spin} = -U\sum_{\substack{ \vc{i}\neq \vc{j}}}J(|\vc{i}-\vc{j}|)\, \hat{\vc{S}}_{\vc{i}}\cdot \hat{\vc{S}}_{\vc{j}}\,
\end{equation}
with couplings $J(\vc{i}-\vc{j})$ defined as 
\begin{equation}
J(\vc{i}-\vc{j}) = \sum_{\vc{l}\alpha} |W_{\alpha}(\vc{l}-\vc{i})|^2|W_{\alpha}(\vc{l}-\vc{j})|^2\,.
\end{equation}

The physical content of the spin Hamiltonian~\eqref{eq:eff_Ham} is that the effective spins located at unit cells $\vc{i}$ and $\vc{j}$ interact through the \textit{density overlap} of Wannier functions centered at unit cells $\vc{i}$ and $\vc{j}$, substantiating the intuition that the density overlap of Wannier functions is responsible for the superconducting order. The ground state of the effective spin Hamiltonian is a product state where all spins are aligned to the same direction. In the language of the attractive Hubbard model this corresponds to the BCS wave function~\eqref{eq:BCS_grand} where the order parameter is given by the expectation value $\Delta_{\vc{i}} = -U \langle \hat{\vc{S}}^-_{\vc{i}} \rangle$. Moreover, the spin Hamiltonian is manifestly $SU(2)$ invariant.

The main drawback of the spin Hamiltonian~(\ref{eq:eff_Ham}) is that it is not gauge invariant. The above truncation depends on the gauge choice for the Wannier states, namely on the definition of the operators $\hat{d}_{\vc{i}\sigma}$. Whereas the BCS ground state and the $SU(2)$ symmetry is correctly reproduced in any gauge, the choice of gauge affects in a substantial way the low-energy spectrum of the effective Hamiltonian. We show in the next section that there is a preferred gauge choice, the maximally localized Wannier functions in the Marzari-Vanderbilt sense.~\cite{Marzari:1997} In this basis the spin Hamiltonian is the best possible approximation of $\overline{\mathcal{H}}_{\rm int}$ in the sense that the gauge invariant result for the superfluid density obtained in Ref.~\onlinecite{Peotta:2015} is recovered.

The subtle point of the gauge noninvariance of the spin Hamiltonian is very well illustrated in the Creutz-Hubbard model. As we mentioned before, the Wannier functions of the Creutz model can be chosen to be perfectly localized on a plaquette, see Fig.~\ref{fig:Creutz}. These plaquette states are in fact the maximally localized Wannier functions in the Marzari-Vanderbilt sense. If $\overline{\mathcal{H}}_{\rm int}$ is expanded in annihilation and creation operators of these plaquette states, pair-breaking and pair-creation terms subject to truncation are absent, and the mapping from the projected interaction Hamiltonian to $\mathcal{\hat H}_{\rm spin}$ is exact. However, this is not the case for arbitrary choices of the Wannier functions. Utilizing the maximally localized lower band Wannier functions of the Creutz model, we obtain the projected Hamiltonian given by

\begin{equation}\label{eq:projected_Ham}
\begin{split}
\overline{\mathcal{H}}_{\rm int}&= -\frac{U}{4}\sum_{\vc{i}} \hat{\rho}_{\vc{i}\uparrow}\hat{\rho}_{\vc{i}\downarrow}-\frac{U}{8}\sum_{\vc{i}}\left(\hat{\rho}_{\vc{i}-1\uparrow}\hat{\rho}_{\vc{i}\downarrow} +\hat{\rho}_{\vc{i}-1\downarrow}\hat{\rho}_{\vc{i}\uparrow}\right ) \\
&-\frac{U}{8}\sum_{\vc{i}}\left(\hat{d}_{\vc{i}-1\uparrow}^\dagger\hat{d}_{\vc{i}-1\downarrow}^\dagger \hat{d}_{\vc{i}\downarrow} \hat{d}_{\vc{i}\uparrow} +\rm{H.c.}\right)\\
&-\frac{U}{8}\sum_{\vc{i}}\left(\hat{d}_{\vc{i}\uparrow}^\dagger \hat{d}_{\vc{i}\downarrow} \hat{d}_{\vc{i}-1\downarrow}^\dagger \hat{d}_{\vc{i}-1\uparrow} +\rm{H.c.}\right)\,,	
\end{split}
\end{equation}
where $\hat{\rho}_{\vc{i}\sigma}=\hat{d}^\dagger_{\vc{i}\sigma}\hat{d}_{\vc{i}\sigma}$.
Indeed, we see that the above Hamiltonian does not mix the subspaces with different number of pairs, and hence for the Creutz model the above described truncation is not an approximation. Using Eqs.~(\ref{eq:def_spin2}), we can map the projected Hamiltonian~(\ref{eq:projected_Ham}) into the ferromagnetic $XXX$ chain with Hamiltonian 
\begin{equation}\label{eq:feroXXX}
	\mathcal{\hat H}_{\rm spin} = -\frac{U}{4}\sum_{ \vc{i}} \hat{\vc{S}}_{\vc{i}}\cdot \hat{\vc{S}}_{\vc{i}+1}\,.
\end{equation}

It is worth mentioning that the spin Hamiltonian~(\ref{eq:eff_Ham}) describes also the completely polarized ferromagnetic state of the corresponding repulsive Hubbard model with half-filled lowest flat band. As discussed in Section~\ref{subsec:model} in the case of the repulsive Hubbard model the kinetic Hamiltonian is spin-isotropic. In this case it is energetically more favorable that all the Wannier states are only singly occupied, i.e., we consider  the subspace defined by $\big(\hat{d}_{\vc{i}\uparrow}^\dagger\hat{d}_{\vc{i}\uparrow} +\hat{d}_{\vc{i}\downarrow}^\dagger\hat{d}_{\vc{i}\downarrow}\big)\ket{\psi}= \ket{\psi}$. Hence, the only remaining degree of freedom is the spin and the effective spin of Eq.~(\ref{eq:eff_Ham}) coincides with the true spin of the fermions. In fact the attractive and repulsive flat band models are exactly related by a particle-hole transformation up to first order in $U/E_{\rm gap}$, while this mapping is broken by interband transitions. The effective spin Hamiltonian~\eqref{eq:eff_Ham} expresses in a concise way the duality between the ferromagnetic and BCS ground states in a flat band. 

\section{Superfluid weight from spin chain}
\label{sec:SFW_spin}
The gauge noninvariance of the spin Hamiltonian~\eqref{eq:eff_Ham} manifests in the fact that the superfluid weight evaluated from it depends on the choice of the Wannier functions. On the other hand, in Ref.~\onlinecite{Peotta:2015} a gauge invariant result for the superfluid weight of a flat band in the presence of an attractive Hubbard interaction has been derived using mean-field BCS theory. The solution of this inconsistency is that, within the spin Hamiltonian approximation, there exists a preferred gauge choice for the Wannier functions which allows to obtain a result for the superfluid weight as close as possible to the gauge invariant result of Ref.~\onlinecite{Peotta:2015}. 
Specifically, in the following we prove the inequality
\begin{equation}\label{eq:trace_inequality}
\mathrm{Tr}D_{\rm s}^{\rm (spin)} \geq \mathrm{Tr}D_{\rm s}
\end{equation}
between the traces of the superfluid weight tensor $D_{\rm s}^{\rm (spin)}$ obtained from the spin Hamiltonian~\eqref{eq:eff_Ham} and the gauge invariant result of Ref.~\onlinecite{Peotta:2015}
\begin{equation}\label{eq:exact_Ds}
[D_{\rm s}]_{i,j} =  \frac{4n_\phi U\nu(1-\nu)}{(2\pi)^d\hbar^2}\int_{\rm B.Z.}d^d\vc{k}\,\mathrm{Re}\, \mathcal{B}_{ij}(\vc{k})\,,
\end{equation}
valid in arbitrary dimension $d$. Here the Quantum Geometric Tensor (QGT) $\mathcal{B}_{ij}(\vc{k})$ is defined by
\begin{equation}\label{eq:QGT}
\mathcal{B}_{ij}(\vc{k}) = 2\bra{\partial_{k_i}g_{\vc{k}}}(1-\ket{g_{\vc{k}}}\bra{g_{\vc{k}}})
 \ket{\partial_{k_j}g_{\vc{k}}}\,,
\end{equation}
where $g_{\vc{k}}(\alpha) = \braket{\alpha}{g_{\vc{k}}}$ are the flat band Bloch functions. Moreover, we show that the gauge noninvariant quantity $\mathrm{Tr}D_{\rm s}^{\rm (spin)}$ attains a global minimum if the maximally localized Wannier functions in the Marzari-Vanderbilt sense are used. In this preferred gauge the superfluid weight tensors calculated using the two different methods coincide in one dimension, while for $d \geq 2$ the spin Hamiltonian generally overestimates the superfluid weight as shown by Eq.~\eqref{eq:trace_inequality}. 

Our argument is based on the fact that the superfluid weight calculated from the spin Hamiltonian is proportional to a functional, called $F_{\rm ov}$, that measures the degree of overlap between  Wannier functions. This functional is similar to the Marzari-Vanderbilt functional, called  $F_{\rm MV}$, which measures the spread of the Wannier functions.~\cite{Marzari:1997} The above results follow from some general relations between these two functionals whose detailed proof is given in the Appendix~\ref{app:functional_relations}. Our proof relies on some assumptions: the first is the uniform pairing condition~\eqref{eq:condition_exact} which is also assumed in Ref.~\onlinecite{Peotta:2015} to derive Eq.~\eqref{eq:exact_Ds}. Since we have proved that the BCS wave function is exact in this case, the use of mean-field BCS theory and thus the result of Eq.~\eqref{eq:exact_Ds} are justified. 
The second assumption is the existence of a gauge in which the Bloch functions are periodic and analytic functions of quasi-momentum, which is equivalent to requiring that the flat band has zero Chern number(s).~\cite{Brouder:2007,Panati:2007, Panati:2013, Monaco:2016} The final assumption is Eq.~\eqref{eq:condition_2} which is discussed in Appendix~\ref{app:functional_relations}. All conditions are verified in the case of the Creutz ladder.

The overlap functional is introduced by considering the state with finite uniform superfluid current given approximately, for small phase-winding wavevector $\vc{q}$, by the ansatz
\begin{equation}\label{eq:BCS_with_sfcurrent}
\ket{\Omega,\vc{q}} = \prod_{\vc{j}}\left(u+e^{2i\vc{q}\cdot\vc{r}_{\vc{j}}}
v\hat{d}^\dagger_{\vc{j}\uparrow}\hat{d}^\dagger_{\vc{j}\downarrow}\right)\ket{\emptyset}\,.
\end{equation}
It is easy to verify that $\bra{\Omega,\vc{q}} \hat{\vc{S}}^{-}_{\vc{j}}\ket{\Omega,\vc{q}} = uve^{2i\vc{q}\cdot\vc{r}_{\vc{j}}} = \sqrt{\nu(1-\nu)}e^{2i\vc{q}\cdot\vc{r}_{\vc{j}}}$ and $\bra{\Omega,\vc{q}} \hat{\vc{S}}^{z}_{\vc{j}}\ket{\Omega,\vc{q}} =  (v^2-u^2)/2 = \nu$.
Therefore, the energy change due to the finite phase-winding with wavevector $\vc{q}$ is given by 
\begin{equation}\label{eq:spin_model_energy}
\begin{split}
\Delta E(\vc{q}) &= \bra{\Omega,\vc{q}} \mathcal{\hat{H}}_{\rm spin}\ket{\Omega,\vc{q}} - \bra{\Omega,\vc{0}} \mathcal{\hat{H}}_{\rm spin}\ket{\Omega,\vc{0}}
\\ &= N_{\rm c}U\nu(1-\nu)\big(F_{\rm ov}(2\vc{q})[W]-F_{\rm ov}(\vc{0})[W]\big)\geq 0\,.
\end{split}
\end{equation}
Here we have defined the \textit{overlap functional} for  Wannier functions $F_{\rm ov}(\vc{q})[W]$ which reads
\begin{equation}\label{eq:ov}
\begin{split}
&F_{\rm ov}(\vc{q})[W] = -\sum_{\vc{i},\vc{j},\alpha}|W_\alpha(\vc{i})|^2\,|W_\alpha(\vc{j})|^2e^{i\vc{q}\cdot(\vc{r}_{\vc{i}}-\vc{r}_{\vc{j}})}\\ &= -\sum_{\vc{i}-\vc{j}}\sum_{\vc{l},\alpha}|W_\alpha(\vc{l}-\vc{i})|^2\,|W_\alpha(\vc{l}-\vc{j})|^2e^{i\vc{q}\cdot(\vc{r}_{\vc{i}}-\vc{r}_{\vc{j}})}\,.
\end{split}
\end{equation}
We call it the overlap functional since, from the second line of the above equation, it is apparent that it can be expressed as the density overlap between the Wannier functions located at unit cells $\vc{i}$ and $\vc{j}$, given by $\sum_{\vc{l},\alpha}|W_\alpha(\vc{l}-\vc{i})|^2|W_\alpha(\vc{l}-\vc{j})|^2$, summed over all possible relative position vectors $\vc{i}-\vc{j}$. 

We also introduce the \textit{Marzari-Vanderbilt localization functional} which differs from the overlap functional only by an additional summation $\sum_{\beta}$ over the orbitals
\begin{equation}\label{eq:MV}
\begin{split}
F_{\rm MV}(\vc{q})[W]
&= - \sum_{\vc{i}\alpha}\sum_{\vc{j}\beta}
|W_\alpha(\vc{i})|^2|W_\beta(\vc{j})|^2 
e^{i\vc{q}\cdot(\vc{r}_{\vc{i}}-\vc{r}_{\vc{j}})}\,.
\end{split}
\end{equation}
Properly speaking, the Marzari-Vanderbilt localization functional
is given by $\nabla_{\vc{q}}^2F_{\rm MV}$. Indeed, one has
\begin{equation}\label{eq:MV_2}
\begin{split}
&\nabla_{\vc{q}}^2F_{\rm MV}(\vc{q}= 0)[W] = 
\sum_{\vc{i}\alpha}\sum_{\vc{j}\beta}|W_\alpha(\vc{i})|^2|W_\beta(\vc{j})|^2(\vc{r}_{\vc{i}}-\vc{r}_{\vc{j}})^2 \\
&= 
2\bigg[\sum_{\vc{i}\alpha}|W_{\alpha}(\vc{i})|^2\vc{r}_{\vc{i}}^2 - \bigg(\sum_{\vc{i}\alpha}|W_{\alpha}(\vc{i})|^2\vc{r}_{\vc{i}}\bigg)^2\bigg]\,.
\end{split}
\end{equation}
Upon replacing the summations with integrals $\sum_{\vc{i}\alpha}\to \int d^d\vc{r}$ in the last equation, the usual definition of the Marzari-Vanderbilt localization functional in the continuum case~\cite{Marzari:2012,Marzari:1997} is recovered. In this sense both $F_{\rm ov}$ and $F_{\rm MV}$ are ``generating functionals" whose expansion in the wavevector $\vc{q}$ generates various moments of the Wannier function density distribution. However, we use the same name for both~\eqref{eq:MV} and~\eqref{eq:MV_2} since the object we are referring to should be clear from the context. The same applies to the overlap functional.

The main technical results of this section are some general relations between the two functionals which are proved in the Appendix~\ref{app:functional_relations} under the conditions stated above. Let us first introduce the maximally localized Wannier functions in the Marzari-Vanderbilt sense, denoted by $\overline{W}_{\alpha}(\vc{i})$, that are the global minimizers of the Marzari-Vanderbilt functional, namely, they satisfy 
\begin{equation}
\nabla_{\vc{q}}^2F_{\rm MV}(\vc{q}= 0)[\overline{W}] \leq \nabla_{\vc{q}}^2F_{\rm MV}(\vc{q}= 0)[W]
\end{equation}
for all Wannier functions $W$ obtained from $\overline{W}$ by a gauge transformation.
The first result is that the maximally localized Wannier functions in the Marzari-Vanderbilt sense are in fact global minimizers of the overlap functional as well; in other words,
\begin{equation}\label{eq:result_1}
\nabla_{\vc{q}}^2F_{\rm ov}(\vc{q}= 0)[\overline{W}] \leq \nabla_{\vc{q}}^2F_{\rm ov}(\vc{q}= 0)[W]\,.
\end{equation}
The second result is the equality, up to the constant factor $n_\phi$ defined in Eq.~\eqref{eq:condition_exact}, of the second derivatives of the two functionals calculated on the maximally localized Wannier functions
\begin{equation}\label{eq:result_2}
\partial_{q_i}\partial_{q_j} F_{\rm ov}(\vc{q}= 0)[\overline{W}] = n_\phi\partial_{q_i}\partial_{q_j} F_{\rm MV}(\vc{q}= 0)[\overline{W}]\,.
\end{equation}
Note that in general the values of the two functionals are different if they are calculated using arbitrary Wannier functions. As consequence of Eqs.~\eqref{eq:spin_model_energy} and \eqref{eq:result_2} one has
\begin{equation}\label{eq:D_spin_MV}
\begin{split}
[D^{\rm (spin)}_{\rm s}]_{i,j} &=  \frac{1}{V\hbar^2} \partial_{q_i}\partial_{q_j} \Delta E(\vc{q}) \\ &= \frac{4n_\phi U\nu(1-\nu)}{V_{\rm c}\hbar^2} \partial_{q_i}\partial_{q_j}F_{\rm MV}(\vc{q}=0)\,,
\end{split}
\end{equation}
where the first equality is the general definition of superfluid weight and the second equality holds only
if the coefficients in the spin Hamiltonian are calculated using maximally localized Wannier functions. For any other gauge one obtains the inequality~\eqref{eq:trace_inequality} as an immediate consequence of Eqs.~\eqref{eq:spin_model_energy},~\eqref{eq:result_1} and~\eqref{eq:D_spin_MV}.

The crucial point of Eq.~\eqref{eq:D_spin_MV} is that the superfluid weight obtained from the spin Hamiltonian has been related to the Marzari-Vanderbilt functional which is well-known to be the sum of a gauge invariant term and a gauge noninvariant one.~\cite{Marzari:1997,Marzari:2012} Not by chance, the result of Eq.~\eqref{eq:exact_Ds} can be recovered from the gauge invariant term of the Marzari-Vanderbilt functional. 
In fact one has $\partial_{q_i}\partial_{q_j}F_{\rm MV}(\vc{q}=0) = \Omega^I_{ij}+\widetilde{\Omega}_{ij}$ where the gauge invariant term is expressed in term of the QGT as 
\begin{equation}\label{eq:gauge_inv_loc_func}
\Omega^I_{ij} = \frac{V_c}{(2\pi)^d}\int_{\rm B.Z.} d^d\vc{k}\, \mathrm{Re}\,\mathcal{B}_{ij}(\vc{k})\,.
\end{equation}
Using Eqs.~\eqref{eq:D_spin_MV} and~\eqref{eq:gauge_inv_loc_func} one recovers Eq.~\eqref{eq:exact_Ds} from the gauge invariant part $\Omega^I_{ij}$ only, which is a positive semidefinite matrix. Also the gauge noninvariant part $\widetilde{\Omega}_{ij}$ is positive semidefinite, which means that the spin Hamiltonian overestimates the superfluid weight of the original Hubbard model if $\widetilde{\Omega}_{ij}\neq 0$. However, in one dimension $\widetilde{\Omega}_{ij}=0$ if maximally localized Wannier functions are used. In general this is not true in $d\geq 2$ due to the noncommutativity of the components of the projected position operator.~\cite{Marzari:1997} Also it can be shown that  $\widetilde{\Omega}_{ij}$ is not vanishing in general when the Berry curvature is nonzero.

A concrete example of the above general results is provided by the Creutz ladder. Indeed, it is easy to verify that the plaquette states are maximally localized Wannier functions in the Marzari-Vanderbilt sense, and in this preferred gauge the number of Cooper pairs is a conserved quantity as shown by Eq.~\eqref{eq:projected_Ham}. As a consequence there is no approximation in going from the projected interaction Hamiltonian $\overline{\mathcal{H}}_{\rm int}$ to the spin Hamiltonian $\mathcal{\hat H}_{\rm spin}$ and the result $D^{\rm (spin)}_{\rm s}$ for the superfluid weight  coincides with Eq.~\eqref{eq:exact_Ds}. On the other hand, if we had chosen to perform the truncation in any other gauge, we would have obtained a superfluid weight which is strictly larger than the correct one as it can be seen from Eq.~\eqref{eq:trace_inequality}  and the results of Appendix~\ref{app:functional_relations}.

We now summarize the results of this section and comment on their significance. The first result is that the approximation underlying the spin effective Hamiltonian~\eqref{eq:eff_Ham} is justified at least in one dimension, since the result of Eq.~\eqref{eq:exact_Ds} is reproduced \textit{exactly}. This means that the spin Hamiltonian is not only able to capture the correct ground state, but also low-lying excited states. In higher dimensions this approximation is justified only if $\widetilde{\Omega}$  is significantly smaller than $\Omega^{I}$ in some sense. In particular, we anticipate that effects due to the Berry curvature, which are absent in one dimension, may play an important role and the low-energy effective Hamiltonian may differ substantially from Eq.~\eqref{eq:eff_Ham}. This is reflected in that we have purposefully avoided the case of nonzero Chern number which is the result of a nonzero average Berry curvature.

As a second important result, we believe we have provided a good understanding of the reason why the quantum metric (the real part of the QGT) enters in the result of Eq.~\eqref{eq:exact_Ds} for the superfluid weight of a flat band. The physical mechanism for transport in a flat band is the correlated hopping of Cooper pairs induced by the Hubbard interaction between Wannier functions with finite density overlap. This picture is nicely captured by the effective Hamiltonian~\eqref{eq:eff_Ham}. We emphasize, though, that the gauge invariant part of the Marzari-Vanderbilt functional measures the spread of the Wannier functions and not their overlap. As we have proved, under some conditions the two distinct functionals measuring spread and overlap of Wannier functions, respectively,  are equivalent, but this may not be true in general. Better insight into this matter can be gained by studying specific models which do not satisfy the conditions required for the validity of the relations~\eqref{eq:result_1} and~\eqref{eq:result_2} between the overlap and spread functionals.

Finally, we have provided a clean example where it is necessary to employ a specific preferred gauge for the Wannier functions in order to obtain a good approximation for the low-energy Hamiltonian. Very often the maximally localized Wannier functions in the Marzari-Vanderbilt sense are used to derive simplified model Hamiltonians from electronic structure calculations. This is a generally accepted heuristic prescription, but we have shown here that there are some rigorous underlying constrains behind it, in our case the minimization of an observable quantity such as the superfluid weight.

\section{Second order corrections to the 
effective Hamiltonian and finite compressibility}
\label{sec:second_order}

Contrary to the repulsive case for which the completely polarized ferromagnetic state is the exact ground state for any value of $U$ if the half-filled flat band is the lowest band, in the attractive case the BCS wave function is the exact ground state of the effective Hamiltonian~\eqref{eq:SW_exp} only up to the first order term. As we shall see now the second order term in the expansion leads to a breaking of the emergent $SU(2)$ symmetry.
It is convenient to study the breaking of the emergent symmetry in the specific case of the Creutz model which is particularly simple. However, this is a general fact, since this symmetry is not present in the generic model under consideration.

The off-diagonal part, as defined in Eq.~\eqref{eq:off-diag-def}, of the interaction term can be written as
\begin{gather}
\mathcal{O}(\mathcal{\hat H}_{\rm int}) = \hat{A}_\uparrow + \hat{A}_\downarrow + \hat{B} + \text{H.c.} \\
\label{eq:A}
\hat{A}_\sigma = -U \sum_{\vc{i}\alpha}\tilde{c}^\dagger_{\vc{i}\alpha\sigma}\bar{c}_{\vc{i}\alpha\sigma} \bar{n}_{\vc{i}\alpha\bar\sigma}\mathcal{\hat{P}}\,,\\
\label{eq:B}
\hat{B} =- U \sum_{\vc{i}\alpha}\tilde{c}^\dagger_{\vc{i}\alpha\uparrow}\bar{c}_{\vc{i}\alpha\uparrow}\tilde{c}^\dagger_{\vc{i}\alpha\downarrow}\bar{c}_{\vc{i}\alpha\downarrow}\mathcal{\hat{P}}\,,
\end{gather}
where $\bar\sigma = \downarrow$ if $\sigma = \uparrow$ and viceversa.
The operators $\hat{A}_\sigma,\, \hat{B}$ introduced above correspond to virtual processes where one particle with spin $\sigma$ or two particles with opposite spins are excited to the upper band, respectively.
Then the second order term in the expansion of the effective Hamiltonian~\eqref{eq:SW_exp} 
reads
\begin{equation}\label{eq:eff_Ham_Creutz}
\mathcal{\hat H}_{\rm eff}^{(2)} = \frac{1}{2}\mathcal{\hat P}[\mathcal{L}(\mathcal{\hat H}_{\rm int}),\mathcal{O}(\mathcal{\hat H}_{\rm int})]\mathcal{\hat P} = -\frac{\hat{A}_\uparrow^\dagger\hat{A}_\uparrow}{E_{\rm g}}-\frac{\hat{A}_\downarrow^\dagger\hat{A}_\downarrow}{E_{\rm g}} -\frac{\hat{B}^\dagger \hat{B}}{2E_{\rm g}}\,.
\end{equation}
In this last result we have made crucial use of the fact that the upper band of the Creutz ladder is also flat. For a generic lattice Hamiltonian with flat bands the computation of the second order term is more involved. 

In order to compute the expansion of the products $\hat{A}_\sigma^\dagger\hat{A}_\sigma,\,\hat{B}^\dagger \hat{B}$ in terms of Wannier operators the following identity is useful
\begin{equation}\label{eq:identity_a}
\mathcal{\hat P}\tilde{c}_{\vc{i}\alpha\sigma}\tilde{c}^\dagger_{\vc{j}\beta\sigma}\mathcal{\hat P} =[\delta_{\alpha\beta}\delta_{\vc{i},\vc{j}}-P^{\sigma}_{\alpha,\beta}(\vc{i}-\vc{j})]\mathcal{\hat P}\,.
\end{equation}
Recall that we impose time-reversal symmetry in the attractive Hubbard model which implies $P^{\uparrow}_{\alpha\beta}(\vc{i}-\vc{j})= [P^{\downarrow}_{\alpha\beta}(\vc{i}-\vc{j})]^* = P_{\alpha\beta}(\vc{i}-\vc{j})$. 
Using Eq.~\eqref{eq:identity_a} we obtain
\begin{widetext}
\begin{equation}\label{eq:A_result}
\begin{split}
\hat{A}_\sigma^\dagger\hat{A}_\sigma&=\frac{U^2}{4}\sum_{\vc{i},\alpha}\bar{n}_{\vc{i}\alpha\uparrow}\bar{n}_{\vc{i}\alpha\downarrow}\mathcal{\hat{P}}-\frac{U^2}{64}\sum_{\vc{i}}[\,\hat{\rho}_{\vc{i}\uparrow}\hat{\rho}_{\vc{i}\downarrow} +\hat{\rho}_{\vc{i}-1\sigma}\hat{\rho}_{\vc{i}-1\bar\sigma}\hat{\rho}_{\vc{i}\bar\sigma}+\hat{\rho}_{\vc{i}-1\bar\sigma}\hat{\rho}_{\vc{i}\sigma}\hat{\rho}_{\vc{i}+1\bar\sigma}+\hat{\rho}_{\vc{i}-1\bar\sigma}\hat{\rho}_{\vc{i}\sigma}\hat{\rho}_{\vc{i}\bar\sigma}\\
&+ \hat{d}^\dagger_{\vc{i}+1\sigma}\hat{d}^\dagger_{\vc{i}+1\bar\sigma}\hat{d}_{\vc{i}-1\bar\sigma}\hat{d}_{\vc{i}-1\sigma}(1-\hat{\rho}_{\vc{i}\bar\sigma})
+ \hat{d}^\dagger_{\vc{i}+1\sigma}\hat{d}^\dagger_{\vc{i}+1\bar\sigma}\hat{d}_{\vc{i}\bar\sigma}\hat{d}_{\vc{i}\sigma}(1+\hat{\rho}_{\vc{i}-1\bar\sigma})
+ \hat{d}^\dagger_{\vc{i}\sigma}\hat{d}^\dagger_{\vc{i}\bar\sigma}\hat{d}_{\vc{i}-1\bar\sigma}\hat{d}_{\vc{i}-1\sigma}(1+\hat{\rho}_{\vc{i}+1\bar\sigma})\\
&-\hat{d}^\dagger_{\vc{i}+1\sigma}\hat{d}_{\vc{i}-1\sigma}\hat{d}^\dagger_{\vc{i}-1\bar\sigma}\hat{d}_{\vc{i}+1\bar\sigma}\hat{\rho}_{\vc{i}\bar\sigma}-\hat{d}^\dagger_{\vc{i}\sigma}\hat{d}_{\vc{i}-1\sigma}\hat{d}^\dagger_{\vc{i}-1\bar\sigma}\hat{d}_{\vc{i}\bar\sigma}\hat{\rho}_{\vc{i}+1\bar\sigma}-\hat{d}^\dagger_{\vc{i}+1\sigma}\hat{d}_{\vc{i}\sigma}\hat{d}^\dagger_{\vc{i}\bar\sigma}\hat{d}_{\vc{i}+1\bar\sigma}\hat{\rho}_{\vc{i}-1\bar\sigma}+\textrm{H.c.}\,]\,,
\end{split}
\end{equation}

\begin{equation}\label{eq:B_result}
\hat{B}^\dagger\hat{B}=\frac{U^2}{4}\sum_{\vc{i},\alpha}\bar{n}_{\vc{i}\alpha\uparrow}\bar{n}_{\vc{i}\alpha\downarrow}\mathcal{\hat{P}}+\frac{U^2}{64}\sum_{\vc{i}}(\,\hat{\rho}_{\vc{i}\uparrow}\hat{\rho}_{\vc{i}\downarrow}+2\hat{d}^\dagger_{\vc{i}\uparrow}\hat{d}^\dagger_{\vc{i}\downarrow}\hat{d}_{\vc{i}-1\downarrow}\hat{d}_{\vc{i}-1\uparrow}+\hat{d}^\dagger_{\vc{i}+1\uparrow}\hat{d}^\dagger_{\vc{i}+1\downarrow}\hat{d}_{\vc{i}-1\downarrow}\hat{d}_{\vc{i}-1\uparrow}+\textrm{H.c.}\,)\,.
\end{equation}
\end{widetext}
Second order processes give rise to three-site terms in the effective Hamiltonian, such as three-site density interaction in the first line of 
Eq.~\eqref{eq:A_result}, correlated pair hopping in the second line of Eq.~\eqref{eq:A_result} and correlated exchange interaction in the third line of Eq.~\eqref{eq:A_result}. It is interesting that even at second order the number of pairs is a conserved quantity, therefore the exchange terms can  dropped in the attractive case and the effective Hamiltonian takes again the form of a spin Hamiltonian with pseudospin operator given by Eq.~\eqref{eq:def_spin2}. We do not know if this is the case even at higher orders and if the number of pairs is an exact conserved quantity of the full model. The breaking of the $SU(2)$ symmetry is evident in $\hat{B}^\dagger\hat{B}$
which produces only two-site operators such as pair hopping. The pair hopping is not compensated by any term of the form $\hat{\rho}_{\vc{i}-1\uparrow}\hat{\rho}_{\vc{i}\downarrow}\,$, and hence the resulting spin chain is an easy-plane ferromagnet. The most important consequence is that the divergent compressibility becomes finite as shown in the following Section. Only the $\hat{B}^\dagger\hat{B}$
term gives rise to a finite compressibility.

\section{Compressibility}
\label{sec:compressibility}

In Section~\ref{sec:exact_BCS} we showed that in the attractive case the BCS state is the exact ground state of the first order term of the effective Hamiltonian~\eqref{eq:SW_exp} with energy $E^{(1)}/L=(2\varepsilon_0-Un_{\phi})\nu$, where $L$ denotes the number of rungs (unit cells) in the ladder, and for the Creutz ladder $n_{\phi}^{-1} = 2$. The second order correction to the ground state energy can be easily obtained by calculating the expectation value $\bra{\Omega} \mathcal{\hat H}_{\rm eff}^{(2)} \ket{\Omega}$, where $\ket{\Omega}$ is given by Eq.~\eqref{eq:BCS_grand}. Below all the expectation values are calculated in this state. It is straightforward to check the following expectation values of single-site operators
\begin{gather}
\langle \hat{\rho}_{\vc{i}\uparrow}\hat{\rho}_{\vc{i}\downarrow} \rangle = \langle \hat{\rho}_{\vc{i}\sigma} \rangle   = v^2 = \nu\,,\\
\langle \hat{d}_{\vc{j}\uparrow}^\dagger\hat{d}_{\vc{j}\downarrow}^\dagger \rangle  = 
\langle \hat{d}_{\vc{j}\downarrow}\hat{d}_{\vc{j}\uparrow}\rangle  = uv = \sqrt{\nu(1-\nu)}\,,\\
\langle\hat{d}_{\vc{j}\sigma}^\dagger\hat{d}_{\vc{j}\bar{\sigma}}\rangle=0\,.
\end{gather}

Due to the fact that the BCS wave function $\ket{\Omega}$ in the grand canonical ensemble is a product state, the expectation values of operators acting on multiple sites can be expressed in terms of the above ones. For example,
\begin{gather}
\langle \hat{\rho}_{\vc{i}-1\uparrow}\hat{\rho}_{\vc{i}\downarrow}\hat{\rho}_{\vc{i}+1\uparrow} \rangle = \langle \hat{\rho}_{\vc{i}-1\uparrow} \rangle \langle \hat{\rho}_{\vc{i}\uparrow} \rangle \langle \hat{\rho}_{\vc{i}+1\uparrow} \rangle = \nu^3\,, \label{eq:exp1}\\
\begin{split}
&\langle \hat{d}^\dagger_{\vc{i}-1\uparrow}\hat{d}^\dagger_{\vc{i}-1\downarrow}
\hat{d}_{\vc{i}\downarrow}\hat{d}_{\vc{i}\uparrow}
\hat{\rho}_{\vc{i}+1\downarrow} \rangle \\ &= 
 \langle\hat{d}^\dagger_{\vc{i}-1\uparrow}\hat{d}^\dagger_{\vc{i}-1\downarrow} \rangle \langle \hat{d}_{\vc{i}\downarrow}\hat{d}_{\vc{i}\uparrow} \rangle \langle \hat{\rho}_{\vc{i}+1\uparrow} \rangle =  \nu^2(1-\nu)\,.
 \end{split} \label{eq:exp2}
\end{gather}
With the above expectation values we are now in a position to evaluate the expectation values of the operators $A^{\dagger}_\sigma A_{\sigma}$ and $B^{\dagger}B$. From Eq.~(\ref{eq:A_result}) we  obtain that $\langle A^{\dagger}_\sigma A_{\sigma}\rangle=0$ and the only nonvanishing contribution comes from Eq.~(\ref{eq:B_result}) and it is given by $\langle B^{\dagger}B\rangle /L=U^2\nu/4-3U^2\nu^2/32$. Hence, from Eq.~(\ref{eq:eff_Ham_Creutz}) the ground state energy of the system up to second order is 
\begin{equation}\label{eq:GS_energy}
\frac{E^{(2)}}{L}=\underbrace{\left(2\varepsilon_0-\frac{U}{2}-\frac{U^2}{8E_{\rm gap}}\right)}_{c_1^{\rm SW}}\nu+\underbrace{\frac{3U^2}{64E_{\rm gap}}}_{c_2^{\rm SW}}\nu^2.	
\end{equation}
The term quadratic in the filling ensures a finite compressibilty since $\kappa^{-1}\propto \nu^2\partial^2E/\partial\nu^2$. Note that the above expression of the ground state energy is valid in the thermodynamic limit, where $L\to \infty$. The coefficients of the linear term and quadratic term in $\nu$ in Eq.~(\ref{eq:GS_energy}) are denoted by $c_1^{\rm SW}$ and $c_2^{\rm SW}$, respectively.

We use DMRG~\cite{White:1992, White:1992a, Schollwock:2005, Schollwock:2011} simulations to find the ground state energy of the original Hubbard model on ladders of length up to $L=26$ with periodic boundary conditions. We typically keep up to $M=2000$ states and check the convergence with respect to $M$. For a fixed length $L$ we measure the ground state energy $E_L(\nu)$ of the system for different densities $\nu$ and by fitting a quadratic function $E_L(\nu)/L = c_1(L)\nu + c_2(L)\nu^2$ we extract the coefficients $c_1(L)$ and $c_2(L)$. Now by applying this procedure for increasing values of $L$, we can see that $c_1(L)$ and $c_2(L)$ have a power law dependence on $1/L$, which allows us to extrapolate this dependence and find the infinite length values $c_1^{\infty}$ and $c_2^{\infty}$. Then these extrapolated values we can compare to the coefficients $c_1^{\rm SW}$ and $c_2^{\rm SW}$ of Eq.~(\ref{eq:GS_energy}) calculated analytically.

\begin{figure}[tb]
\centering	\includegraphics{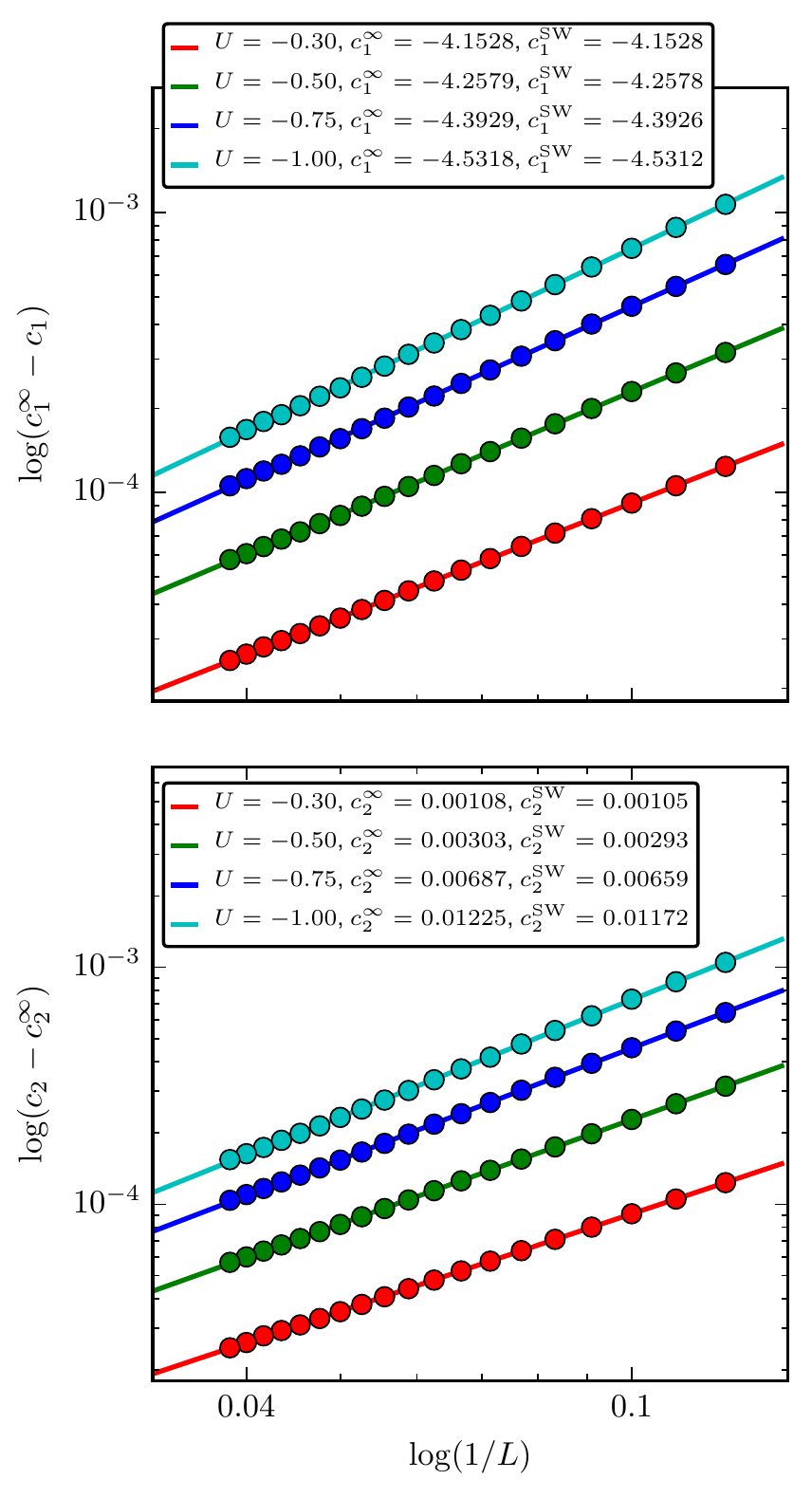}
\caption{\label{fig:log-log}(Color online) The power law dependence of $c_{1}(L)$ (top plot) and $c_{2}(L)$ (bottom plot) on $1/L$ plotted in log-log scale. The dots represent the DMRG results the lines correspond to fitted function. Different colors correspond to different values of the interaction $U$. Here, we set $t=1$. In the legends the values of $c_{1(2)}^{\infty}$ and $c_{1(2)}^{\rm SW}$ are presented for direct comparison.} 
\end{figure}

In Fig.~\ref{fig:log-log} we plot in log-log scale the difference of $c_{1(2)}(L)$ and the extrapolated infinite system size value $c_{1(2)}^{\infty}$ versus $1/L$. These plots clearly show that we have a power law dependence. Moreover, in the legends we give a direct comparison of the extrapolated infinite system size values of the coefficients $c_{1(2)}^{\infty}$ with the values $c_{1(2)}^{\rm SW}$ obtained from effective model. From these numbers one can see that the effective model and the DMRG simulations with high accuracy give the same energy dependence. With the increase of the interactions strength $U$, the small difference between the values of $c_{2}^{\infty}$ and $c_{2}^{\rm SW}$ grows approximately from $2.5\%$ (for $-U=-0.3$) to $4.5\%$ (for $-U=-1.0$). This difference is a result of higher order interband effects, which become more relevant for a large interaction strength $U$.
\section{Relation between winding number and superfluid weight}
\label{sec:winding}

Using Eq.~(\ref{eq:exact_Ds}) and the positive semidefinitness of the QGT it has been shown in Ref.~\onlinecite{Peotta:2015} that in two dimensions a flat band with nonzero Chern number is guaranteed have a finite superfluid weight with the bound $D_s\geq |\mathcal{C}|$, in appropriate units. In this Section we derive a similar bound in one dimension and show that a flat band with non-zero \textit{winding number} has finite Drude weight $D$. In one dimension the superfluid weight is not defined and it is better to use the Drude weight instead, which measure the ability of the system to sustain ballistic transport. The Drude weight and the superfluid weight are equal in gapped systems in $d\geq 2$. 

Let us consider a general one-dimensional system with chiral symmetry represented by a unitary operator $\Gamma$, which anticommutes with the Bloch Hamiltonian $\{\Gamma,\mathcal{H}(k)\}=0$. The presence of a gap in the band structure allows to deform the Bloch Hamiltonian $\mathcal{H}(k)$ adiabatically into a flat band Hamiltonian defined as $h(k)=\mathcal{Q}_k-\mathcal{P}_k$, where $\mathcal{P}_k$ and $\mathcal{Q}_k=1-\mathcal{P}_k$ are the projectors to the subspace of the bands below and above the gap, respectively. Then a topological invariant for the system known as the winding number is given by~\cite{Wen:1989,Sato:2009} 
\begin{align}
\mathcal{W}=\frac{i}{4\pi}\int_{\rm B.Z.} dk \Tr\left[\Gamma h(k)\partial_kh(k)\right]\\=\frac{i}{\pi}\int_{\rm B.Z.} dk \Tr\left(\Gamma\mathcal{P}_k\partial_k\mathcal{P}_k\right)\,,\label{eq:winding_num}
\end{align}
where the last equality was obtained by using $h(k)=1-2\mathcal{P}_k$  and the fact that the integral of the full derivative term $\partial_k\left[\Tr\left(\Gamma\mathcal{P}_k\right)\right]$ over the Brillouin zone vanishes. 

The QGT defined in Eq.~(\ref{eq:QGT})  can be written alternatively as $\mathcal{B}_{ij}(\vc{k}) = 2\Tr\left[\mathcal{P}_{\vc{k}}\partial_{k_i}\mathcal{P}_{\vc{k}}\partial_{k_j}\mathcal{P}_{\vc{k}}\right]$ and in 1D it becomes a real valued scalar function $\mathcal{B}(k)=2\Tr\left[\mathcal{P}_k\left(\partial_k\mathcal{P}_k\right)^2\right]$. Using the cyclicity of the trace, the properties $\mathcal{P}_k^2=\mathcal{P}_k$ and $\mathcal{P}_k^{\dagger}=\mathcal{P}_k$ of the projector and by inserting $\Gamma\Gamma^\dagger$ one can easily verify that 
$\mathcal{B}(k)=2\Tr\left[\mathcal{P}_{k}(\partial_{k}\mathcal{P}_{k})^2\right]=2\Tr[A_k^{\dagger}A_k]$ with $A_k=\Gamma\mathcal{P}_k\partial_k\mathcal{P}_k$. Note that the integrand of Eq.~(\ref{eq:winding_num}) is equal to $\Tr(A_k)$.

Let $\mathcal{M}$ be the integral of the QGT over the Brillouin zone defined by
\begin{equation}\label{eq:def_M}
	\mathcal{M}=\frac{1}{a\pi}\int_{\rm B.Z.}dk \,\mathcal{B}(k)\,, 
\end{equation}
where the factor $1/a\pi$ is for convenience and $a$ is the lattice constant. Now we can prove the following lower bound for $\mathcal{M}$
\begin{equation}
\begin{split}
\mathcal{M}&=\frac{2 }{a\pi} \int_{\rm B.Z.}dk\,\Tr(A_k^\dagger A_k)\\ &\geq \frac{2}{a\pi\rk(A_k)}\int_{\rm B.Z.}dk\,\left|\Tr(A_k)\right|^2 \\ &\geq \frac{1}{\rk(A_k)\,\pi^2}\left|\int_{\rm B.Z.}dk\Tr\left(\Gamma\mathcal{P}_k\partial_k\mathcal{P}_k\right)\right|^2\geq \frac{\mathcal{W}^2}{\rk(\mathcal{P}_k)}\,.\label{eq:sc_bound}
\end{split}
\end{equation}
Above in the first step we have used the fact that for any $n\times n$ matrix $A$ the inequality $\Tr(A^{\dagger}A)\geq |\Tr(A)|^2/\rk(A)$ holds,~\cite{Wolkowicz:1980} where $\rk(A)$ denotes the rank, and then we have used a special case of Schwartz's inequality. In the last step we have used the definition of the winding number given in Eq.~(\ref{eq:winding_num}) and that $\rk(A_k)\leq \rk(\mathcal{P}_k)$. Note that $\rk(\mathcal{P}_k)$ is equal to the number of bands below the gap and obviously does not depend on $k$.
 
Since the Drude weight $D$ is given by Eq.~\eqref{eq:exact_Ds} in any dimension, including $d=1$ where the superfluid weight is not defined, we conclude that  in the flat band limit the lower bound $D\propto \mathcal{M} \geq \mathcal{W}^2/\rk(\mathcal{P}_k)$ holds. This bound implies that in 1D a topological nontrivial flat band ($\mathcal{W}\neq 0$) has a guaranteed finite Drude weight for finite attractive interaction.

For the Creutz ladder the chiral symmetry is represented by the Pauli matrix $\sigma_2$. Since there is only one band below the gap we have $\rk(\mathcal{P}_{k,{\rm C}})=1$. This means that according to above bound we have $\mathcal{M}_{\rm C}\geq (\mathcal{W}_{\rm C})^2$. Using the Bloch functions in Eq.~\eqref{eq:plaquette_Bloch}, one can easily check that $\mathcal{B}_{\rm C}(k)=a^2/2$ and $\Tr\left(\sigma_2\mathcal{P}_{k,{\rm C}}\partial_k\mathcal{P}_{k,{\rm C}}\right)=-ia/2$. Form Eq.~(\ref{eq:winding_num}) we obtain $\mathcal{W_{\rm C}}=1$ and from Eq.~(\ref{eq:def_M}) we get $\mathcal{M}_{\rm C}$ = 1. Hence, for the Creutz ladder the actual value of $\mathcal{M}$ coincides with the lower bound, and we have that the inequality is saturated, i.e., $\mathcal{M}_{\rm C}=(\mathcal{W}_{\rm C})^2$.

\section{Discussion and conclusion}

In this work we have provided some definite results that help in characterizing the low-energy physics of attractive Hubbard models in flat bands. Some results are also relevant for the repulsive case at half-filling in the context of flat band ferromagnetism. 
The first important conclusion is that under the condition of uniform pairing in the flat band orbitals given in Eq.~\eqref{eq:condition_exact} the BCS wave function is the exact zero temperature ground state at the level of the projected interaction Hamiltonian. This result justifies the use of mean-field theory in Ref.~\onlinecite{Peotta:2015} where the relation~\eqref{eq:exact_Ds} between superfluid weight and QGT has been derived and extends similar results valid only for bipartite lattices~\cite{Julku:2016} and Landau levels.~\cite{Peotta:2015} It is an interesting question to understand what happens if the uniform pairing condition is violated. In fact in many flat band models this condition is satisfied due to symmetry reasons, such as in the Creutz ladder, the Lieb lattice, Kagome lattice and Landau levels, but it is also easy to construct realistic models where it is not. 

We note that it is trivial to prove that the ferromagnetic ground state is exact in a repulsive spin-isotropic Hubbard model if the flat band is half-filled and is the lowest lying band. Our result is more subtle since the BCS ground state is not exact anymore if higher order terms in the SW expansion are included and it is not the exact ground state of the full Hamiltonian, even if the flat band is the lowest lying band. Indeed,  at first order in $U$, the attractive and repulsive cases are related by a duality that takes the form of a particle-hole transformation. This duality is broken if interband transitions are taken into account, as we show explicitly by computing the second order term in the SW expansion for the effective Hamiltonian.

We show that the asymptotic exactness (for small $U$) of the BCS wave function is the manifestation of on emergent $SU(2)$ symmetry and give the corresponding generators. This $SU(2)$ symmetry corresponds to the $SU(2)$ spin symmetry in the  repulsive case and implies that the compressibility of the system is diverging since the ground state energy is linear in the flat band filling $\nu$. This is very useful result in the context of ultracold gas experiments since the compressibility is routinely measured. 

Due to the breaking of the emergent $SU(2)$ symmetry by interband transitions, the compressibility is in fact finite and positive as we show explicitly in the case of the Creutz-Hubbard model, by calculating the ground state energy up to the second order term in $U$. Indeed, the compressibility \textit{decreases} if the coupling constant of the attractive interaction decreases $-U \to -\infty$. In many other fermionic system the behavior is rather different since the compressibility \textit{increases} if the interaction becomes more attractive. It would be interesting to understand if this persists at nonzero temperature in order to make comparison with ultracold gas experiments.

A byproduct of our proof of the asymptotic exactness of the BCS wave function is the relation $n_\phi\langle \bar{n}_{\vc{i}\alpha\uparrow} \rangle = n_\phi\langle\bar{n}_{\vc{i}\alpha\downarrow} \rangle= \langle \bar{n}_{\vc{i}\alpha\uparrow}
\bar{n}_{\vc{i}\alpha\downarrow}\rangle$ between local density and double occupancy and $n_\phi$ is the parameter defined in Eq.~\eqref{eq:condition_exact}. This result can be tested in ultracold gas experiments, for example, with site-resolved imaging, and should be valid even at quite high temperatures.

We have shown how from the projected interaction Hamiltonian $\overline{\mathcal{H}}_{\rm kin}$ many terms can be dropped resulting in an effective ferromagnetic Heisenberg $\mathcal{\hat{H}}_{\rm spin}$ model which is a very convenient approximation computationally and offers an intuitive model of a flat band superfluid as a ferromagnet.
 This approximation is justified since many exact properties are preserved such as the exactness of the BCS wave function and the emergent $SU(2)$ symmetry, while the major drawback is that the invariance under gauge transformations of the Wannier functions is broken. We provide at least a partial solution to this problem by noticing that in the preferred gauge of maximally localized Wannier functions the gauge invariant result for the superfluid weight of Ref.~\onlinecite{Peotta:2015} is recovered in one dimension. In higher dimensions the result is in general larger than the gauge invariant one of Ref.~\onlinecite{Peotta:2015} since the gauge noninvariant term $\widetilde{\Omega}$ of the Marzari-Vanderbilt localization functional is nonzero even when calculated on the maximally localized Wannier functions. It has been recently proved that the Marzari-Vanderbilt localization functional is finite if and only if the band has zero Chern number.~\cite{Monaco:2016} This means that the approximation leading to the spin Hamiltonian in Eq.~\eqref{eq:eff_Ham} fails in the case of bands with nonzero Chern number, a case which we have not discussed in this work and is a very interesting subject for future research. Also the case of a flat band with zero Chern number but large Berry curvature is interesting since in this case $\widetilde{\Omega}$ is large as well and the spin Hamiltonian~\eqref{eq:eff_Ham} is probably not a very good approximation.

Our results are relevant for the general problem of estimating the critical temperature of the superconducting transition. Notice that the BCS wave function in a flat band is identical to the one in the strong coupling limit $U/t \to +\infty$ ($t$ is the hopping energy scale)
with the difference that in the flat band the Cooper pair wave function coincides with the the Wannier function and is delocalized on $n_{\phi}^{-1}$ orbitals, while   
in the strong coupling limit the Cooper pair is localized on a single orbital. Indeed, it is well known that the strong coupling limit can be treated with the SW transformation as an expansion in the parameter $t/U$ which is essentially the inverse of the parameter $U/E_{\rm gap}$ employed here. Similarly to the strong coupling limit, the critical temperature is controlled by the collective fluctuations and the relevant energy scale is $\sim \hbar^2 D_{\rm s}$. In the flat band one has $\hbar^2 D_{\rm s} \propto U$, while in the strong coupling limit $\hbar^2 D_{\rm s} \propto t^2/U$. The effective Hamiltonian at the leading order is in both cases an isotropic Heisenberg ferromagnet with the difference that in the strong coupling limit the range of the ferromagnetic couplings coincides with the range of the hopping terms in the kinetic Hamiltonian while they can be long-ranged in the   
flat band limit. If the flat band is intersected by other bands, as in the case of the Lieb lattice~\cite{Julku:2016} and kagome lattice,~\cite{Huber:2010} or has nonzero Chern number, the Wannier functions are algebraically decaying to the extent that the Mermin-Wagner theorem may not be applicable and long range order can appear in dimension two even in the presence of a continuous symmetry. These considerations give an idea of the subtleties involved in providing reliable estimates of the critical temperature in flat band systems.

Some of the results provided here are of more general interest, beyond the topic of flat band superconductivity. The same effective spin Hamiltonian~\eqref{eq:eff_Ham} is the first order term in the SW expansion for a spin-isotropic repulsive Hubbard model with half-filled flat band, and is therefore the low-energy theory of a  flat band ferromagnet. In particular, this  shows that the QGT plays a crucial role also for flat band ferromagnets, in which case the quantity corresponding to the superfluid weight is the spin stiffness. This has not been realized so far and it is an interesting topic for further investigations. Other results of general interest are the inequality $\mathcal{M} \geq \mathcal{W}^2$ between the quantum metric and the winding number, and the various relations between the overlap functional and the Marzari-Vanderbilt localization functional, which may find applications in other contexts. In particular, we believe that inequalities between quantum metric and the rich variety of topological invariants known so far can be found, similarly to what has been done here for the one-dimensional winding number and in previous works for the Chern number. 

A promising way to verify our results would be via transport measurements in cold atom experiments, as pioneered in Ref.~\onlinecite{Brantut:2012}. Furthermore, for a tangible experimental realization of the Creutz ladder, we refer the reader to Refs.~\onlinecite{Mazza:2012, Mugel:2016} and the recent advances in implementing artificial magnetic fields for neutral atoms in synthetic dimensions.~\cite{Aidelsburger:2013, Miyake:2013, Celi:2014, Mancini:2015, Stuhl:2015}

\acknowledgments
\noindent We acknowledge useful discussions with Long Liang, Arun Paramekanti and Jacques Tempere. S.P. thanks the authors of Ref.~\onlinecite{Monaco:2016} for sharing their results before publication. Our DMRG simulations are based on the ALPS libraries.~\cite{ALPS2:2011,Dolfi:2014} This work was supported by the Academy of Finland through its Centers of Excellence Programme  (2012-2017) and under Project Nos. 263347, 251748, 272490, and 284621 and
by the European Research Council (ERC-2013-AdG-340748-CODE). This project has received funding from the European Union's Horizon 2020 research and innovation programme under the Marie Sklodowska-Curie grant agreement No. 702281 (FLATOPS). M.T. and S.D.H. acknowledge support from the Swiss National Science Foundation. S.P. thanks the Pauli Center for Theoretical Studies, and M.T. thanks the COMP Center of Excellence of Aalto University for their hospitality.

\appendix
\section{Relations between the overlap and Marzari-Vanderbilt functionals}
\label{app:functional_relations}

Before providing a proof of Eqs.~\eqref{eq:result_1} and~\eqref{eq:result_2}, we present other relations  between the the two functionals for completeness.
The two functionals can be identified up to the constant factor $n_\phi$ for arbitrary values of $\vc{q}$ if the condition
\begin{equation}\label{eq:Wannier_property}
|W_\alpha(\vc{i})|^2 = |W_\beta(\vc{i})|^2 \quad \text{for every} \quad \alpha,\beta,\vc{i}\,,
\end{equation}
is satisfied. For example, the above equation holds if the orbital $W_\alpha(\vc{i})$ is the plaquette state of the Creutz ladder. However, even in the Creutz ladder this condition does not hold for arbitrary gauge choices, and therefore is not useful for our purposes.
On the other hand by using the elementary inequality $\sum_{\alpha=1}^n|v_\alpha|^2/n \geq |\sum_{\alpha=1}^nv_\alpha/n|^2$ it is easily shown that 
\begin{equation}\label{eq:func_rel}
F_{\rm ov}(\vc{q})[W] \leq n_\phi F_{\rm MV}(\vc{q})[W]\,.
\end{equation}
Here $n_\phi^{-1}$ is simply the number of $\alpha$'s
for which $W_\alpha(\vc{i})\neq 0$ and is not defined by the condition~\eqref{eq:condition_exact}, which is not necessary for the above inequality to hold. Eq.~\eqref{eq:func_rel} provides a strong inequality but in the wrong sense and  again it is not useful in this context.

The proof of Eqs.~\eqref{eq:result_1} and~\eqref{eq:result_2} relies on the following relation between Wannier functions and Bloch functions
\begin{equation}
\label{eq:identity_WB_bis}
\sum_{\vc{i}}|W_\alpha(\vc{i})|^2e^{i\vc{q}\cdot\vc{r}_{\vc{i}}}
= \frac{V_{\rm c}}{(2\pi)^d}\int_{\rm B.Z.} d^d\vc{k}\,
 g_{\vc{k}+\vc{q}}^*(\alpha)
g_{\vc{k}}(\alpha)\,,
\end{equation}
which is readily derived from the definition of Wannier functions given in Eq.~\eqref{eq:Wannier_def}. Accordingly, we write $F_{\rm ov}(\vc{q})[g]$ and $F_{\rm ov}(\vc{q})[g]$ to the denote that the these are functionals of the Bloch functions $g_{\vc{k}}(\alpha)$ corresponding to the Wannier functions $W_{\alpha}(\vc{i})$. Moreover, we denote by $\bar{g}_{\vc{k}}(\alpha)$ the Bloch functions corresponding to the maximally localized Wannier functions in the Marzari-Vanderbilt sense $\overline{W}_{\alpha}(\vc{i})$.

In the following we assume that the Bloch functions are analytic and periodic functions of the quasi-momentum $\vc{k}$. This allows in particular to perform integration by parts and drop full derivatives which integrate to zero over the whole Brillouin zone. In general it is not possible to find an analytic and periodic gauge if the Chern numbers in $d\geq 2$ are nonzero,~\cite{Brouder:2007,Monaco:2016} while in $d = 1$ this is always possible.~\cite{Kohn:1973,Nenciu:1998} For the Creutz model the smooth and periodic Bloch functions are given in Eq.~\eqref{eq:plaquette_Bloch}. In what follows we restrict to the case of bands with zero Chern numbers. 

Let us first insert Eq.~\eqref{eq:identity_WB_bis} into the definition of the overlap and Marzari-Vanderbilt localization functionals given in Eqs.~\eqref{eq:ov} and~\eqref{eq:MV}, respectively,  and take the Laplacian $\nabla^2_{\vc{q}}$. After integration by parts we are left with
\begin{widetext}
\begin{gather}
\label{eq:F_O_Bloch}
\begin{split}
&\nabla^2_{\vc{q}}F_{\rm ov}(\vc{q}=0)[g] = \frac{2V_{\rm c}}{(2\pi)^d}
\sum_\alpha P_{\alpha\alpha}(\vc{0})\int_{\rm B.Z.} d^d\vc{k}\, \bm\nabla g_{\vc{k}}^*(\alpha)\cdot \bm\nabla g_{\vc{k}}(\alpha)-2\sum_\alpha|\bm\theta_{\alpha}|^2\,,
\end{split}\\
\label{eq:F_MV_Bloch}
\begin{split}
&\nabla^2_{\vc{q}}F_{\rm MV}(\vc{q}=0)[g] = \frac{2V_{\rm c}}{(2\pi)^d} 
\int_{\rm B.Z.} d^d\vc{k}\, \bra{\bm\nabla g_{\vc{k}}}\cdot\ket{\bm\nabla g_{\vc{k}}}
-2|\bm\theta|^2\,.
\end{split}
\end{gather}
\end{widetext}
The second terms on the right hand sides of Eqs.~\eqref{eq:F_O_Bloch} and~\eqref{eq:F_MV_Bloch} depend on the quantities
\begin{gather}
\label{eq:Zak_phase_orbital}
\bm\theta_{\alpha} = -i\frac{V_c}{(2\pi)^d}\int_{\rm B.Z.} d^d\vc{k}\,g_{\vc{k}}^*(\alpha)\bm\nabla g_{\vc{k}}(\alpha)\,,\\
\label{eq:Zak_phase}
\bm\theta = -i\frac{V_c}{(2\pi)^d}\int_{\rm B.Z.} d^d\vc{k}\,\braket{g_{\vc{k}}}{\bm\nabla g_{\vc{k}}} = \frac{V_c}{(2\pi)^d}\int_{\rm B.Z.} d^d\vc{k}\, \vc{A}(\vc{k})\,.
\end{gather}
Eq.~\eqref{eq:Zak_phase} is known as the Zak phase (up to normalization) while the $\bm\theta_\alpha$ in Eq.~\eqref{eq:Zak_phase_orbital} are orbital-resolved Zak phases whose sum give the Zak phase $\bm\theta = \sum_\alpha \bm\theta_{\alpha}$.
The Zak phase can be expressed in terms of the Berry connection $\vc{A}(\vc{k}) = -i\braket{g_{\vc{k}}}{\bm\nabla g_{\vc{k}}}$ as shown in Eq.~\eqref{eq:Zak_phase}. The Zak phase is gauge invariant up to shifts by an arbitrary lattice vector. Indeed, a generic gauge transformation $g_{\vc{k}}\to e^{i\phi_{\vc{k}}}g_{\vc{k}}$ that preserves periodicity and analyticity is given by a phase of the form $\phi_{\vc{k}} = \vc{k}\cdot\vc{r}_{\vc{l}} + \bar\phi_{\vc{k}}$, where $\vc{r}_{\vc{l}}$ is an arbitrary lattice vector and $\bar\phi_{\vc{k}}$ a periodic analytic function, namely $\bar\phi_{\vc{k}} = \bar\phi_{\vc{k}+\vc{G}_j}$ for all reciprocal lattice vectors $\vc{G}_j$. The nonperiodic part of the phase $\phi_{\vc{k}}$ corresponds to a rigid translation of the Wannier function by the lattice vector $\vc{r}_{\vc{l}}$. Under this transformation the Zak phase $\bm{\theta}$ transforms as 
$\bm{\theta} \to \vc{r}_{\vc{l}} + \bm{\theta}$. Instead the orbital-resolved Zak phases $\bm\theta_\alpha$ depend on the choice of gauge for the Bloch functions.

We now restrict ourselves to the case where Eq.~\eqref{eq:condition_exact} holds and let us further assume that
\begin{equation}\label{eq:condition_2}
\overline{\bm{\theta}}_\alpha = \overline{\bm{\theta}}_\beta\quad \text{for}\quad \alpha,\beta \in \mathcal{S}\,,
\end{equation}
when the orbital-resolved Zak phases are calculated on the maximally localized Wannier/Bloch functions, namely $\overline{\bm\theta}_{\alpha} = -i\frac{V_c}{(2\pi)^d}\int_{\rm B.Z.} d^d\vc{k}\,\overline{g}_{\vc{k}}^*(\alpha)\bm\nabla \overline{g}_{\vc{k}}(\alpha)$. This is the case of the Creutz ladder where $\overline{\theta}_{\alpha=1} = \overline{\theta}_{\alpha=2} = a/4$. Eq.~\eqref{eq:condition_2} is the second requirement mentioned previously that allows to derive Eqs.~\eqref{eq:result_1}-\eqref{eq:result_2}. Indeed, by inserting $P_{\alpha\alpha}(\vc{0}) = n_\phi$ and $\bm\theta_{\alpha} = n_\phi\bm\theta$ into Eq.~\eqref{eq:F_O_Bloch} and comparing with Eq.~\eqref{eq:F_MV_Bloch}, one immediately obtains
\begin{equation}
\nabla^2_{\vc{q}} F_{\rm ov}(\vc{q}= 0)[\overline{W}] = n_\phi\nabla^2_{\vc{q}} F_{\rm MV}(\vc{q}= 0)[\overline{W}]\,.
\end{equation}
The same argument can be straightforwardly adapted to prove Eq.~\eqref{eq:result_2}. To prove that the Wannier functions are global minimizers of the overlap functional let us add and subtract a term $n_\phi\frac{2V_c}{(2\pi)^d}\int_{\rm B.Z.} d^d\vc{k}\,|\vc{A}(\vc{k})|^2$
in Eq.~\eqref{eq:F_O_Bloch} and write arbitrary Bloch functions as $g_{\vc{k}} = e^{i\phi_{\vc{k}}}\overline{g}_{\vc{k}}\,$. This leads to 
\begin{widetext}
\begin{equation}\label{eq:inter_step}
\begin{split}
\nabla^2_{\vc{q}}F_{\rm ov}(\vc{q}=0)[g] &= n_\phi\frac{V_c}{(2\pi)^d}\int_{\rm B.Z.} d^d\vc{k}\, \mathrm{Tr}\,\mathcal{B}(\vc{k}) + n_\phi\frac{2V_c}{(2\pi)^d}\int_{\rm B.Z.} d^d\vc{k}\,|\vc{A}(\vc{k})|^2 - 2\sum_\alpha|\bm\theta_\alpha|^2 \\ &= n_\phi\nabla^2_{\vc{q}}F_{\rm MV}(\vc{q}=0)[\overline{g}] + n_\phi\frac{2V_c}{(2\pi)^d}\int_{\rm B.Z.} d^d\vc{k}\,|\bm\nabla\phi_{\vc{k}}|^2 - 2\sum_\alpha\left(\frac{V_c}{(2\pi)^d}\int_{\rm B.Z.} d^d\vc{k}\, |g_{\vc{k}}(\alpha)|^2\bm\nabla\phi_{\vc{k}}\right)^2\,.
\end{split}
\end{equation}

In order to derive the above equation we have used again the conditions in Eq.~\eqref{eq:condition_exact} and Eq.~\eqref{eq:condition_2} and the fact that the Berry connection calculated on the maximally localized Bloch functions is a divergence-free vector field $\bm\nabla\cdot \overline{\vc{A}}(\vc{k}) =  0\,$, as it can be easily shown by varying the Marzari-Vanderbilt functional. Note that the above functional is quadratic in $\bm\nabla\phi_{\vc{k}}$. Moreover, using the inequality
\begin{equation}\begin{split}
\frac{V_c}{(2\pi)^d}\int_{\rm B.Z.} &d^d\vc{k}\, \frac{|g_{\vc{k}}(\alpha)|^2}{n_\phi} |f(\vc{k})|^2
\geq \left|\frac{V_c}{(2\pi)^d}\int_{\rm B.Z.} d^d\vc{k}\, \frac{|g_{\vc{k}}(\alpha)|^2}{n_\phi} f(\vc{k}) \right|^2
\end{split}
\end{equation}
valid for arbitrary functions $f(\vc{k})$, one can see from Eq.~\eqref{eq:inter_step} that the maximaly localized Wannier functions in the Marzari-Vanderbilt sense correspond to a global minimum of the overlap functional, and thus Eq.~\eqref{eq:result_1} is proved.
\end{widetext}
\bibliography{References}
\end{document}